\documentclass[preprint,showpacs]{revtex4}
\usepackage[dvips]{graphicx}

\begin{document}

\title{Spin-isospin nuclear response 
using the existing microscopic Skyrme functionals}
\author{S. Fracasso and G. Col\`o}

\affiliation{Dipartmento di Fisica, Universit\`a degli Studi and INFN,
Sezione di Milano, 20133 Milano, Italy
}

\begin{abstract}
Our paper aims at providing an answer to the question whether one 
can reliably describe the properties of the most important 
spin-isospin nuclear
excitations, by using the available non-relativistic Skyrme energy 
functionals. Our method, which has been introduced in a previous
publication devoted to the Isobaric Analog states, is the 
self-consistent Quasiparticle Random Phase Approximation (QRPA).
The inclusion of pairing is instrumental for describing a 
number of experimentally measured 
spherical systems which are characterized by open
shells. We discuss the effect of isoscalar and isovector pairing
correlations. Based on the results for the Gamow-Teller resonance
in $^{90}$Zr, in $^{208}$Pb and in few Sn isotopes, we draw
definite conclusions on the performance of different Skyrme
parametrizations, and we suggest improvements for future fits. We also
use the spin-dipole resonance as a benchmark of our statements. 
\end{abstract}

\maketitle

\section{Introduction}
\label{intro}

The special role played by the spin-isospin modes for the 
detailed understanding of the structure of nuclei has been 
pointed out since several decades. The subject has been treated 
in review papers~\cite{Osterfeld} and textbooks~\cite{HarakehVanDerWoude}. 
Spin-isospin transitions can occur spontaneously, in the case of 
$\beta$-decay. The simplest case is that of the Gamow-Teller (GT) 
transitions, whose corresponding operator is
\begin{equation}
\vec O_{GT\pm} = \sum_{i=1}^A \vec \sigma(i) t_\pm(i).
\end{equation}
This operator is associated with a model-independent sum rule, 
namely $m_0 \equiv m_0(t_-)-m_0(t_+)=3(N-Z)$ 
where $m_0(t_\pm)$ is the total strength in the given channel. 
Since the early work of K. Ikeda {\em et al.} where this sum rule 
has been introduced~\cite{Ikeda}, it has been clear that in the limited 
energy window accessible to the $\beta$-decay only a limited fraction of 
this sum rule can be found. A collective state should be expected at 
higher energy, and this Gamow-Teller resonance (GTR) has been indeed 
detected in (p,n) experiments starting from the mid-seventies~\cite{Doering}. 
Later, systematic ($^3$He,t) experiments which are characterized by 
much better energy resolution have also been performed. We 
should remind that in nuclei having neutron excess the Ikeda sum rule 
is exhausted almost entirely by states in the $t_-$ channel 
as the Pauli principle hinders the $t_+$ excitations. 

In medium-heavy nuclei, ranging from $^{90}$Zr to $^{208}$Pb, 
the GTR is located somewhat above the Isobaric Analog Resonance 
(IAR) which is also well known from (p,n) and ($^3$He,t) 
experiments. This corresponds to the typical energy region of 
the giant resonances, that is, 10-20 MeV (we refer here to energies 
with respect to the ground state of the mother, or target, nucleus). At the 
same time, the main GT peak(s) turns out to exhaust only about 
50\% of the Ikeda sum rule in these medium-heavy nuclei; 
this percentage becomes about 70\% if the whole strength in the 
neighboring energy region (i.e., below $\approx$ 20 MeV 
in the daughter, or final, nucleus) is 
collected~\cite{RapaportGaarde}. 

The extraction of the strength from the measured cross 
sections is far from being straightforward. However, due 
to their $\Delta L$=0 character, the GTR and IAR angular 
distributions are strongly peaked at 0$^\circ$, and an 
approximate proportionality between the zero-degree cross 
section and the strength has been found under the hypothesis 
of high incident energy, zero momentum transfer and 
neglect of the non-central components of the projectile-target 
interaction~\cite{GoodmanTaddeucci}. 

The problem of the so-called ``missing GT strengh'' has 
considerably attracted the attention of nuclear physicists. 
Some theorists have speculated that part of the missing GT 
strength should be found at very high excitation 
energy ($\approx$ 300 MeV) due to the coupling with the 
internal 1$^+$ excitation of the nucleon, i.e., the 
$\Delta$-isobar (1232 MeV): the reader can consult the 
references quoted in~\cite{Osterfeld}. In other 
calculations~\cite{2p2h}, it has been shown that the usual 
coupling of the one particle-one hole (1p-1h) configurations 
involved in the GTR with two particle-two hole (2p-2h) configurations 
is able to shift strength outside the range accessible to 
experiments and explain in a more conventional fashion 
the missing strength. Experimentally, from the 
multipole-decomposition analysis (MDA) of the cross 
sections measured in the $^{90}$Zr(p,n) experiment at
$E_{\rm p}$=295 MeV~\cite{Wakasa}, it has been argued that 90\% 
of the GT strength can be recovered below 50 MeV excitation
energy, leaving little room for the coupling with the 
$\Delta$-isobar. However, part of the analysis (for instance, 
the estimate of the isovector monopole contribution) has 
been somehow questioned. 

The coupling of simple 1p-1h configurations with more 
complex one, and the high-lying GT strength, are not the 
issue of the present paper. Using the Skyrme Hamiltonian, 
the GTR in $^{208}$Pb has been calculated, beyond simple 
RPA, taking into account the coupling with the continuum 
as well as with configurations made up with a p-h pair 
coupled with a collective vibration~\cite{Col94}. This
calculation has been able to reproduce the values of
the branching ratios associated with the proton decay
of the GTR; at the same time, it has been shown that 
the position of the main GT peak does not change too 
much with respect to simple RPA. 
In Figs.~4 and 5 of Ref.~\cite{Col94} one
can see that the peak is indeed shifted downwards by few hundreds 
of keV. The calculations reported in Ref.~\cite{Dang} 
(also based on the coupling with phonons) 
are much more phenomenological, but the result is similar.
The redistribution of strength mentioned in the
previous paragraph is instead quite sizeable and
this point should be kept in mind for the following
discussion. No such complete 
and fully microscopic calculation, at the level
of four-quasiparticle coupling, is available for the
charge-exchange modes in open shell systems. We 
still need, and this is our first aim here, 
to assess in a clear way the properties of the Skyrme
functionals, complemented by an effective pairing
force, in the spin-isospin channel by studying 
the corresponding excitations within the self-consistent
mean-field framework. As self-consistent calculations, 
we mean Quasiparticle 
Random Phase Approximation (QRPA) calculations based on a 
Hartree-Fock plus Bardeen-Cooper-Schrieffer (HF plus BCS) 
description of the ground state. Our model has been introduced 
and applied to the IAR in Ref.~\cite{Fracasso}. Some
rather preliminary results using the same model have been
presented in Conference proceedings~\cite{Fracasso_previous}.

There are not many self-consistent QRPA calculations available. 
The proton-neutron QRPA based on Skyrme forces in the particle-hole 
(p-h) channel (with a simplified form, i.e., with a separable
approximation), and on the use of a constant pairing gap in HF-BCS plus 
a free residual particle-particle (p-p) interaction, has been
intensively applied to the study of both spherical and
deformed nuclei~\cite{madrid_papers}. The issue is to know 
to what extent instrinsic deformations affect the measured
$\beta$-decay spectra and the authors of~\cite{madrid_papers}
have explored many
isotopic chains, including heavy ones~\cite{madrid_last}.
Later, the first attempt to implement a self-consistent
QRPA scheme based on HFB has been made in Ref.~\cite{Engel}, 
which is another work devoted to $\beta$-decay (in this
case limited to spherical isotopes -- lying 
on the $r$-process nucleosynthesis path). The same group
has studied the high-lying GTR, and the behavior of
different Skyrme parameter sets~\cite{Bender}: we will
discuss in detail, in what follows, the comparison of that 
work with the present one. The charge-exchange modes 
have also been attacked by using
relativistic charge-exchange RPA and QRPA~\cite{rel_arg,
rel_paar,rel_ma,finelli}. 

But, aside from the mentioned ones, 
most of the QRPA calculations are not 
self-consistent. To study the GTR in $^{208}$Pb, the 
quasiparticle-phonon model has been employed in 
Ref.~\cite{Kuzmin}, and the so-called Pyatov method 
in Ref.~\cite{Babacan}. Most of the systematic calculations
(done also for open-shell nuclei and/or for the $\beta$-decay)
are rather based on some empirical mean field 
(e.g., Woods-Saxon) and residual interaction depending, 
in the spin-isospin channel, on a parameter $g_0^\prime$. 
A schematic model of this type can certainly be useful in many 
respects. As we discuss below, predictions of the schematic RPA model 
based on these simple phenomenological ingredients can be regarded 
as a guideline while understanding our results. 
However, we stick on the idea of a full microscopic approach. 
This is of special interest nowadays: if new radioactive 
beam facilities aim at studying spin-isospin properties 
of the exotic systems, constraining this channel in the microscopic 
Hamiltonian must be envisaged, while sticking on phenomenological
inputs may be not appropriate. 

The study of spin-isospin excitations is not only of interest for nuclear structure, 
but also for particle physics or astrophysics. In fact, 
the detailed knowledge of spin-isospin nuclear matrix 
elements (with $\approx$ 20\% accuracy) is required to 
extract from the $\beta\beta$-decay experimental findings 
the hierarchy of neutrino masses. And, in the astrophysical 
sector, the details of the $r$-process nucleosynthesis 
can be understood, once more, only if nuclear masses, photonuclear 
cross sections and $\beta$-decay probabilities are precisely known. 
Last but not least, we mention the importance of knowing 
the neutrino-nucleus interactions in different contexts 
(from the stellar environment, to the case of materials
which are used for crucial experiments on the neutrinos). 
All these motivations lie at the basis of the recent works 
concerning the spin-isospin nuclear modes. 

In the present work, as compared with 
Ref.~\cite{Bender}, we re-discuss 
in particular the role 
of the so-called spin-gradient (or $J^2$) terms of the 
Skyrme energy functionals, and we find somewhat different
results for the GT strength distributions. Moreover, 
we perform a more general analysis since we also  
study the role of the pairing residual interaction (which 
has been neglected in~\cite{Bender}), and we 
devote some attention to the case of another 
kind of spin-isospin excitation, namely the 
isovector spin-dipole 
resonance.   

The isovector spin-dipole (IVSD) resonance is excited 
by the operator 
\begin{equation}
O_{IVSD\pm,J^\pi} = \sum_{i=1}^A r_i 
\left[ \vec Y_{1}(\hat r_i) \otimes 
\vec \sigma(i)\right]_{J^\pi M} t_\pm(i),
\end{equation}
where $J^\pi=0^-,1^-,2^-$. 
The charge-exchange experimental measurements, whether (p,n) or
($^3$He,t), show indeed evidence of $L\neq 0$ strength. Most of this
strength is very fragmented, and an unambigous signature 
for the different multipoles (monopole, dipole etc.) is still 
missing. In theoretical calculations, the spin-dipole
distributions look quite broad, also because of the
presence of three $J^\pi$ components. Some calculations 
for magic nuclei have been available for long time: the
reader can refer to the phenomenological calculations of
Ref.~\cite{Bertsch81} or to the HF plus continuum-RPA of 
Ref.~\cite{auerbach_spin}. Recently, there has 
been new interest in the study of this channel: some 
low-lying transitions which are important for the
$\beta\beta$-decay have in fact first-forbidden character, 
and the reliability of theoretical models in predicting 
properties of $L\neq 0$ charge-exchange transitions is
under discussion. Moreover, it has been suggested that
the precise determination of the IVSD sum rule 
(analogous to the Ikeda sum rule) can be a 
unique probe of the neutron skins, as it is proportional
to $N \langle r^2 \rangle_n - Z \langle r^2 
\rangle_p$~\cite{Krasznahorkay99}. Since it would be highly 
desirable to extract the key parameters governing the 
asymmetry part of the nuclear equation of state 
from the difference of the neutron and proton radii, 
and the experimental determination of neutron radii
by means of scattering data is not very accurate, 
this alternative way of extracting the same quantity
is potentially of great interest (see also~\cite{sag_tbp}).
In the spirit of the present investigation, 
it is important of course to establish whether the 
conclusions about the robustness of the Skyrme-QRPA with
given parameter sets, remain valid when another
multipolarity is studied. 

The outline of our paper is the following. We first
provide the basic information about our formalism 
in Sec.~\ref{forma}, by limiting ourselves to what 
is essential for understanding the rest of the discussion.
One part of the Skyrme functionals that we employ here, 
namely that associated with the so-called $J^2$ terms, 
has been discussed recently, also in 
Ref.~\cite{Bender}; for this reason, we discuss at length
the point of view emerging from our calculations and 
results in Sec.~\ref{J2_treat}. We can then analyze 
the results for the GT and IVSD strength distributions, 
respectively, in Secs.~\ref{GT_res} and~\ref{IVSD_res}, 
and draw relevant conclusions on the performances of the
exisiting Skyrme sets as well as make suggestions for the
future fits. Considerations on the pairing correlations
are made in Sec.~\ref{pairing}, before coming to
the overall conclusions of Sec.~\ref{conclu}. 

\section{Formalism}
\label{forma}

Our model has been introduced in Ref.~\cite{Fracasso} and we will 
focus here only on those aspects which are important for the
understanding of our results. 
We start by dealing with the HF-BCS coupled problem, that is, at 
each iteration we solve in real space the HF equations and the
BCS gap and number equations. For $^{90}$Zr and 
$^{208}$Pb pairing is neglected. For the Sn isotopes, 
the pairing window is the 50-82 neutron shell, and the 
pairing force is the same which has been fitted in 
Ref.~\cite{Fracasso}, namely a zero-range, density-dependent 
interaction of the type
\begin{equation}
V=V_0 \left( 1- \left( {\rho \left( {\vec r_1+\vec r_2\over 2} \right)
\over \rho_{\rm C}} \right)^\gamma \right)\cdot\delta(\vec r_1-\vec r_2),
\label{ppforce}
\end{equation}
with $V_0$=680 MeV$\cdot$fm$^3$, $\rho_{\rm C}$=0.16 fm$^{-3}$ and $\gamma$=1.
It has been checked that when this pairing force is used in connection 
with different Skyrme forces (we consider in this work the 
parameter sets SIII~\cite{Beiner}, SGII~\cite{VanGiai}, 
SLy5~\cite{Chabanat} and SkO$^\prime$~\cite{SkOp}), the 
resulting pairing gaps do not vary too much along the Sn 
isotope chain. 

All the states at positive energy (either those in the BCS pairing 
window or those outside this window, which have occupation factors
$v^2$ equal to zero) are calculated using box boundary conditions: 
that is, our continuum is discretized. Two quasiparticle configurations 
(or particle-hole, in the cases in which pairing is absent) with 
proper $J^\pi$ are built and the QRPA matrix equations, 
\begin{equation}
\left( \begin{array}{cc} A & B \\ -B & -A \end{array} \right) 
\left( \begin{array}{c} X^{(n)} \\ Y^{(n)} \end{array} \right) =  E_n 
\left( \begin{array}{c} X^{(n)} \\ Y^{(n)} \end{array} \right),
\label{matrix}
\end{equation}
are solved in this model space. The upper limit for the 
configurations is chosen so that the results are stable against
variations and the proper sum rules, which are
expected to hold in full self-consistent calculations, 
are indeed exhausted with high accuracy. In the charge-exchange
case, it is known that these sum rules are 
the difference of the non energy-weighted sum rules in 
the two isospin channels $m_0 \equiv m_0(t_-)-m_0(t_+)$, and 
the sum of the energy-weighted sum rules 
$m_1 \equiv m_1(t_-)+m_1(t_+)$. 
The analytic values of these sum rules in the case of 
the Skyrme forces can be found, e.g., in Ref.~\cite{auerbach}. 

In the p-h channel, for the charge-exchange modes, 
the residual interaction reads
\begin{eqnarray}
v^{ph}(\textbf{r}_1,\textbf{r}_2)&=&\delta(\textbf{r}_1-\textbf{r}_2)
\left[v_{01}(r)+v_{11}(r)+v_{01}'+v_{11}'+v_{1}^{(s.o.)}
\right].
\end{eqnarray}
In this formula, the two indices for each of the first four terms 
in square brakets refer 
to the projection in a given $\sigma\tau$-channel. The terms with (without) 
a prime are those which are (are not) velocity-dependent. The last term is 
the isovector part of the spin-orbit residual interaction. 

In the following of this work, our considerations will 
focus on the spin-isospin terms of the p-h residual interaction, the
spin-independent terms being far from dominant or even
not active. For the sake of completeness, we provide
anyway the detailed expressions of all terms:
\begin{eqnarray}
v_{01}(r)&=&
2C_1^{\rho}\left[\rho(r)\right] 
{\vec\tau}_1{\vec\tau}_2,\nonumber\\
v_{11}(r)&=&
2C_1^{S}\left[\rho(r)\right]{\vec\sigma}_1{\vec\sigma}_2
{\vec\tau}_1{\vec\tau}_2,\nonumber\\
v_{01}'&=&
\left[\left(k'^2+k^2\right)\frac{1}{2}\left(C_1^{\tau}-4C_1^{\bigtriangleup 
\rho}\right)
+k'k\left(3C_1^{\tau}+4C_1^{\bigtriangleup \rho}\right)\right]
{\vec\tau}_1{\vec\tau}_2,\nonumber\\
v_{11}'&=&
\left[\left(k'^2+k^2\right)\frac{1}{2}\left(C_1^{T}-4C_1^{\bigtriangleup S}
\right) 
+k'k\left(3C_1^{T}+4C_1^{\bigtriangleup S}\right)\right]{\vec\sigma}_1
{\vec\sigma}_2{\vec\tau}_1{\vec\tau}_2,\nonumber\\
v_{1}^{(s.o.)}&=&
-2iC_1^{\bigtriangledown J}\left(\sigma_1+\sigma_2\right)k' \times k
{\vec\tau}_1{\vec\tau}_2. 
\label{ph_int}
\end{eqnarray}
We remind that 
\begin{eqnarray}
k'&=&-\frac{1}{2i}(\bigtriangledown_1'-\bigtriangledown_2'),\nonumber\\
k &=& \frac{1}{2i}(\bigtriangledown_1-\bigtriangledown_2),
\end{eqnarray}
with the operators acting at right (left) in the case of $k$ ($k'$). 
The parameters entering the above expressions can be written in terms
of those of the Skyrme force. For the convenience of the reader, this
correspondence is explicitly provided in the Appendix. 

In the p-p channel, we fix self-consistently the residual 
isovector pairing
force by exploiting the isospin symmetry, that is, we take 
the isovector proton-neutron pairing interaction to be the same
as the neutron-neutron one used in the BCS description of the 
ground state. The proton-neutron isoscalar pairing cannot
be constrained: presently, we miss a clear indication from 
empirical data about the parameter of the isoscalar pairing
force. Using a quite conservative approach, we present in the 
following results which, unless otherwise stated, correspond
to an isoscalar pairing force equal to the isovector one.
We have tried to give some indication about the sensitivity 
of our results when in the isoscalar channel a strength 
$V_0^{(T=0)}$ different from $V_0^{(T=1)}$ is adopted.
This kind of study has been done in connection with the 
RMF analysis of the charge-exchange modes. No such analysis
has been available so far in the case of the Skyrme 
calculations; we have found results which are
to some extent consistent with those associated with the
RMF study. In that case, a finite-range Gogny pairing force is
employed, but this does not seem to produce macrocopic 
differences with respect to the use of zero-range effective
pairing forces.

\section{Treatment of the $J^2$ terms of the energy functional}
\label{J2_treat}

As mentioned above, the energy functional
includes the so-called spin-gradient, or $J^2$, terms which
are built on the spin-orbit densities. They arise from 
the exchange part of the central Skyrme 
interaction~\cite{VautherinBrink}. The spin-orbit densities 
vanish in the ground state of spin-saturated nuclei but 
they provide a contribution to the spin and spin-isospin parts
of the residual p-h interaction. In the past (with some
exceptions), the $J^2$ terms have been
neglected when fitting the Skyrme parameters; some more
recent parametrizations include them, and in particular
we will consider in the following the sets 
SLy5 and SkO$^\prime$. In the
discussion below, for the sake of simplicity, we will
call type I-forces the Skyrme sets which do not include
the $J^2$ terms in the fit, and type II-forces those 
which do include them. To our knowledge, there is not
a clear indication emerging from the nuclear phenomenology 
whether these $J^2$ terms must be included in a physically
sound energy functional. Of course, if the functional is
derived from a two-body force of the Skyrme type, which
has a momentum dependence, it looks questionable to
drop the $J^2$ terms. We should also notice that the 
$J^2$ terms neither are hard to evaluate, nor they are 
time-consuming if the HF calculation is peformed in 
coordinate space as in the present case. 

In the past many RPA calculations have been performed with 
type I-forces. The authors of~\cite{Bender} have pointed out 
that those calculations (e.g., those of Ref.~\cite{VanGiai}) 
do not respect the full self-consistency, since the 
contributions from the $J^2$ terms are included in the
residual interaction but not in the mean 
field (the authors of~\cite{VanGiai} have also
neglected the spin-orbit residual interaction but this
has practically no effect). 
To respect the Galilean invariance, 
the authors of~\cite{Bender}, when employing 
type I-forces in their work, have adopted the prescription of 
removing from the residual interaction not only the
contribution from the $J^2$ term, but also that from 
the so-called $S\cdot T$ term in the functional. 
This amounts to setting $C_1^T$ equal to
zero (cf. Eq. (\ref{ph_int})) and leads to a substantial
quenching of the velocity-dependent part in 
the spin-isospin channel since $C_1^{\bigtriangleup S}$ 
is not as large. Therefore, we deem that the 
issue should be further discussed here. 

We start from the fact that fitting the Skyrme
parameters is usually done by using, in addition to 
nuclear (or neutron) matter quantities, binding energies 
and charge radii of few selected isotopes (with the 
spin-orbit strength $W_0$ separately adjusted). In $^{208}$Pb, 
the binding energy (charge radius) changes by 0.22\% 
(0.15\%), using the force SLy4, when the $J^2$ terms are 
omitted or inserted. These variations are too
small to allow a clear statement about the manifestation 
of the $J^2$ terms in the benchmarks used for the fit because 
it must be noted that in the protocol for the parameter
fitting presented in Ref.~\cite{Chabanat}, larger 
errors on binding energies and charge radii are imposed 
in the $\chi^2$-formula to let the fit converge (we mean here, larger than
the experimental error bars and larger than the $\approx$ 
0.1-0.2\% variations we just mentioned). Even in $^{120}$Sn, 
which is not used for the parameter fitting but is studied
in the present paper, we find a similar pattern.

\begin{figure}[hbt]
\includegraphics[width=2.8in]{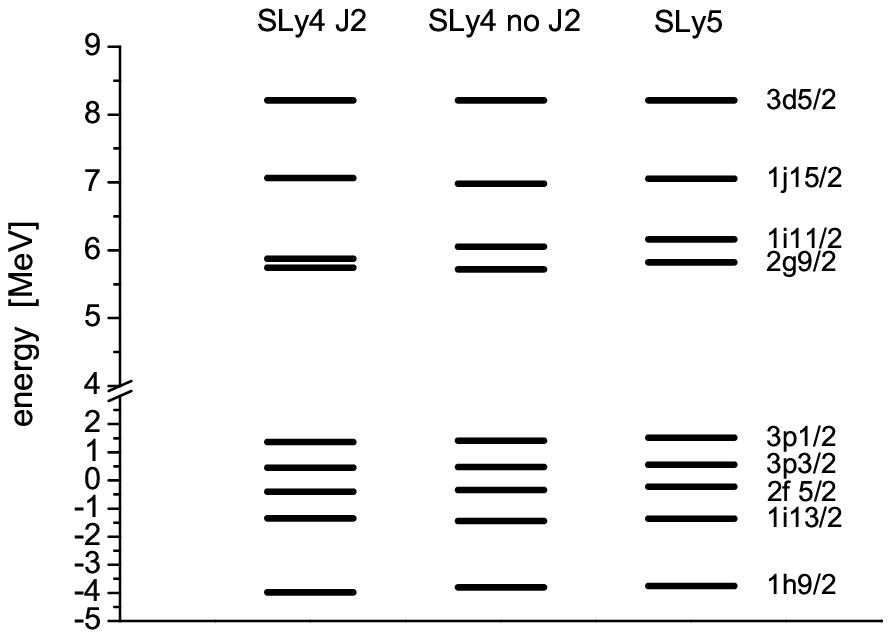}
\includegraphics[width=2.8in]{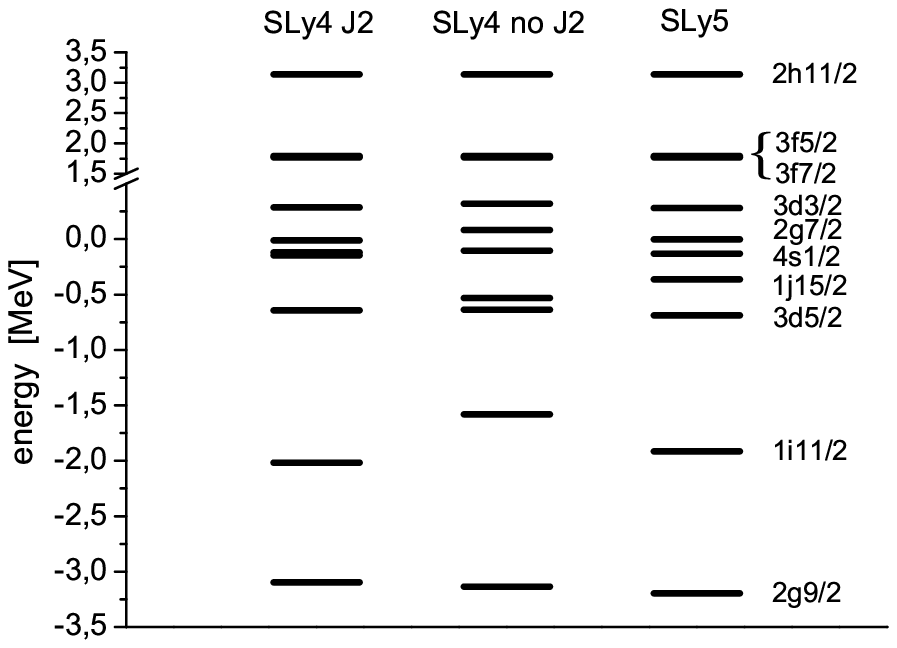}
\vspace{0.3cm}
\includegraphics[width=2.8in]{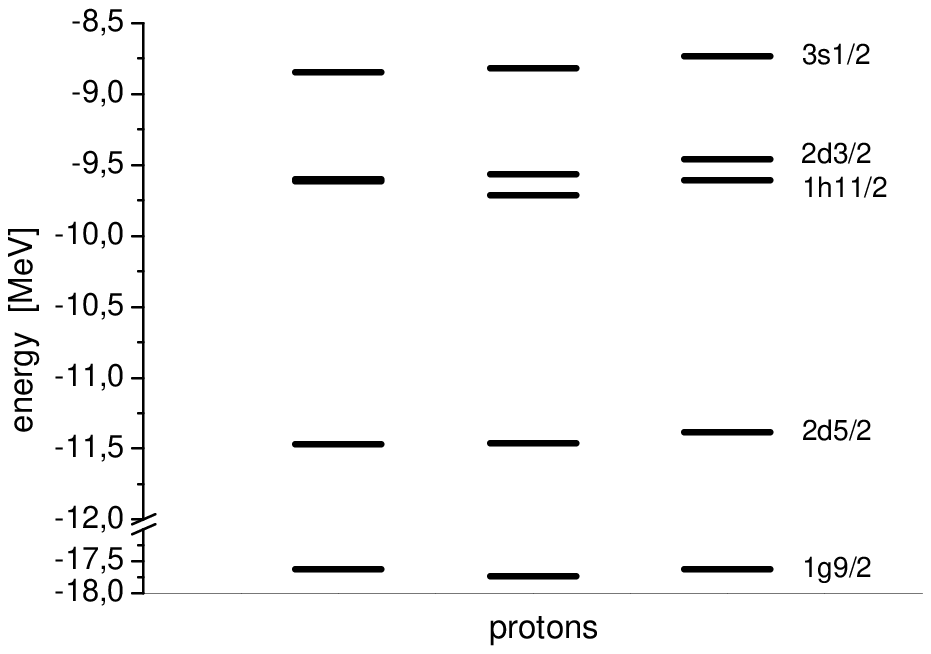}
\includegraphics[width=2.8in]{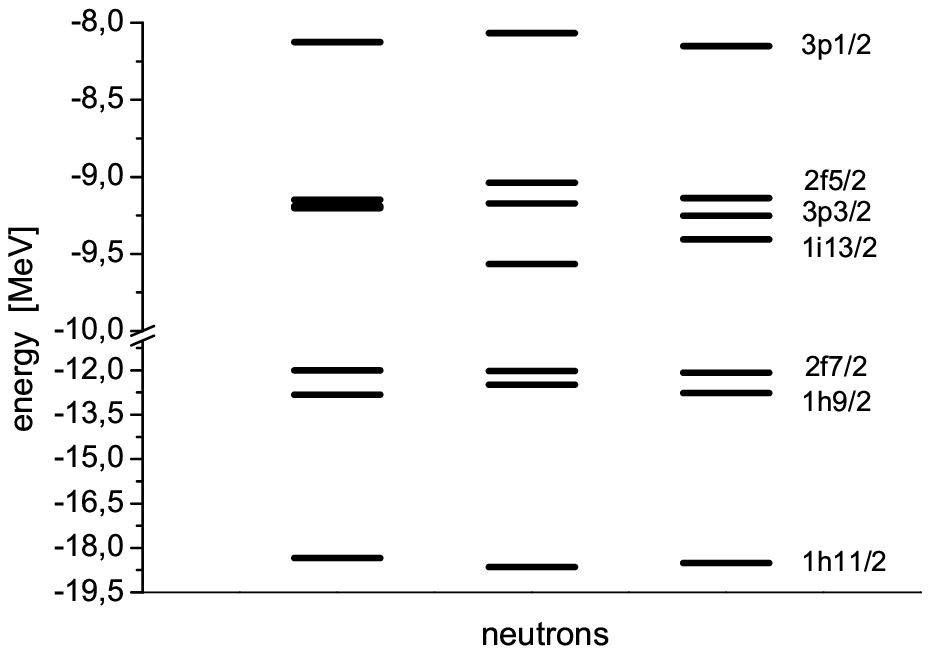}
\caption{\label{fig:levels} Single-particle levels of 
$^{208}$Pb, calculated using Skyrme-HF and employing respectively 
the parametrization SLy4 with and without the $J^2$ terms and
the parametrization SLy5. The left (right) part refers to 
protons (neutrons). The lower (upper) panel includes levels 
below (above) the Fermi energy. 
}
\end{figure}

\begin{figure}[hbt]
\includegraphics[width=5.5in]{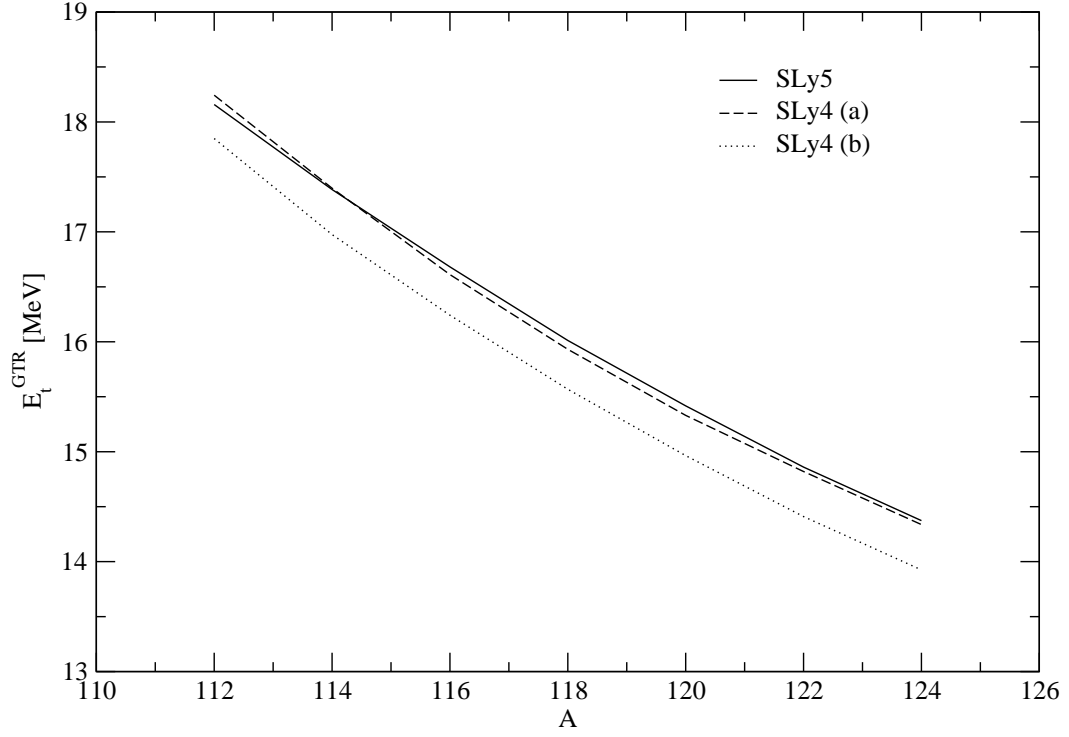}
\caption{\label{fig:J2}
GTR peak energies, along the Sn isotope chain, calculated
by using either the force SLy5 (full line) or the force
SLy4 with (dotted line) 
and without (dashed line) the contribution associated with
the $J^2$ terms in the mean field.
See the text for the discussion.
}
\end{figure}

On the other hand, the effect of the $J^2$ terms on the 
single-particle spectrum becomes appreciable. In 
Fig.~\ref{fig:levels} 
we display the highest occupied and lowest unoccupied 
proton and neutron levels in $^{208}$Pb. With few 
exceptions the spectra
of SLy4 and SLy5 are similar, the proton (neutron) 
levels being in general slightly lower (higher) in energy
in the case of SLy4. The spectrum
associated with SLy4 plus the $J^2$ terms is instead somewhat
different: in fact, one notices that the $j_>$ ($j_<$) 
spin-orbit partners are raised (lowered) in energy, both for
protons and neutrons, up to 400 keV. Accordingly, the 
unperturbed energies of the $j_> \rightarrow j_<$ configurations
are reduced. The net overall effect is that 
the main GTR peak varies only by 60 keV between
SLy4 and SLy5, but it varies by 0.5 MeV when the $J^2$ terms
are added to the SLy4 mean field. This is due to the fact that only 
$j_> \rightarrow j_<$ configurations are present in the
GTR wavefunction calculated with the Lyon parameter sets. 
We show the variation of the GTR peak energy along the 
Sn isotopes in Fig.~\ref{fig:J2}. 
A stronger effect of the 
$J^2$ terms is that associated with the spin-isospin residual 
p-h force. In fact, if we remove the part corresponding to 
the $J^2$ term in the energy functional from the p-h 
interaction, the GTR peak energy changes by about 2 MeV.
Qualitatively similar conclusions can be deduced from
the study of the Sn isotopes. 

This detailed study led us to the following conclusions. 
The $J^2$ terms do not manifest themselves so much in the ground state
observables used for the fit of the Skyrme parameters, but 
they do affect some other properties of the nuclear ground state 
like the spin-orbit splitting. Moreover, they play a major role
when GT calculations are performed, mainly because 
of their contribution to the p-h interaction.
Looking at our results, we 
believe that the most natural and physical choice is to omit 
the contribution of the $J^2$ terms 
when calculating the ground state with type-I forces, 
but retain the corresponding contribution in the
residual p-h force. 
In the case of nuclei  
which are not spin-saturated, 
we agree with the authors of~\cite{Bender} 
that this choice breaks self-consistency. If one 
insists on self-consistency, the choice of inserting the 
$J^2$ contribution in the ground state alters the GTR 
energy by about 0.5 MeV, whereas the alternative choice 
of neglecting the $J^2$ contribution systematically 
appears to be quite unnatural. After all, we definitely 
suggest that fits of new Skyrme parameters
are systematically done by inserting the $J^2$ terms. 

We conclude this Section by mentioning that in the
recent literature there have been claims about the
necessity of complementing the usual Skyrme forces 
with tensor terms (even and odd). Together with other
collaborators, the authors of the present paper 
have shown that the contribution of the tensor
effective force can remedy serious and qualitative 
discrepancies between the single-particle levels 
predicted within the Skyrme framework and those 
which are experimentally observed~\cite{tensor}.
Similar discussions can be found in~\cite{tensor2}.
The reason for mentioning this here, is that the 
two-body zero-range tensor force gives the same kind 
of contribution to the mean field of even-even nuclei 
nuclei as the $J^2$ terms. Consequently, 
the tensor force will affect the GT centroid 
energy, and we have estimated its impact by using
sum rule arguments in~\cite{tensor}. Since the aim of
this work is the discussion of the performance 
of the existing functionals we do not come back
to this point in the following. If a new general fit
of Skyrme functionals plus tensor contribution is 
made, and the corresponding (Q)RPA becomes available,
new steps can be undertaken.  

\section{Results for the Gamow-Teller response}
\label{GT_res}

\begin{figure}[hbt]
\includegraphics[width=5.5in]{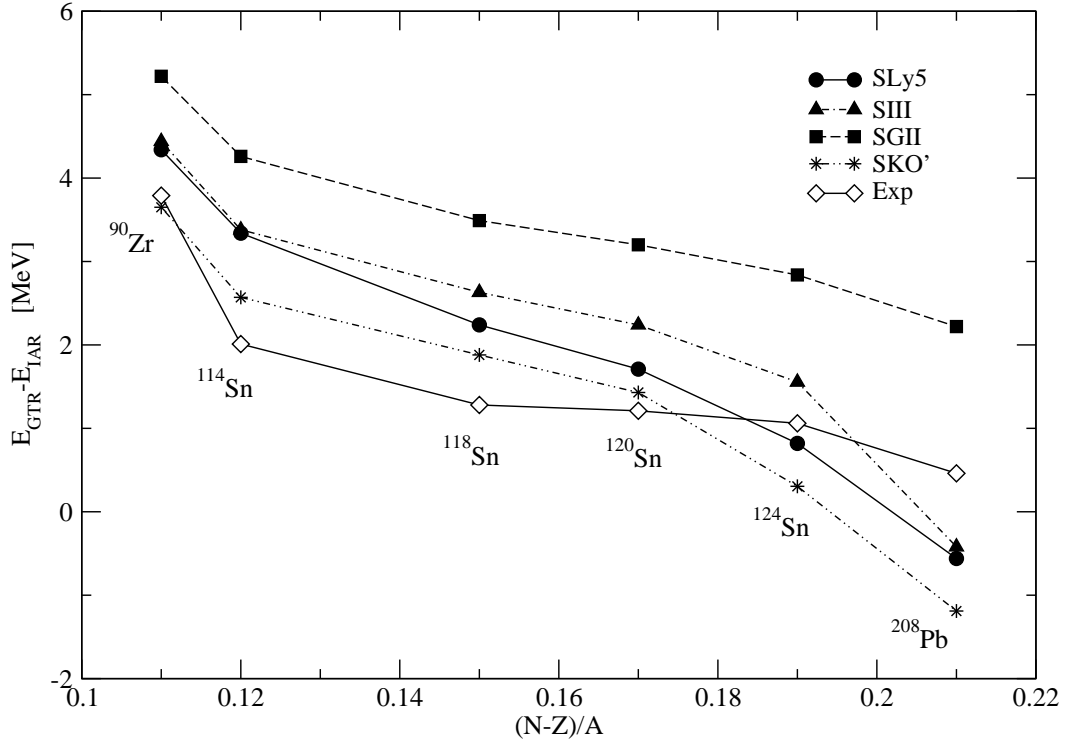}
\caption{\label{fig:trendGT}
Difference between the GTR and the IAR energies in some 
selected spherical nuclei as a function of $(N-Z)/A$. 
Theoretical results associated with different Skyrme
parametrizations are compared with experimental values 
from Refs.~\cite{Pham95,Akimune95,Krasznahorkay01}. 
The related discussion, including details on how the 
energies have been defined, can be found in the text. 
}
\end{figure}

As stated in the Introduction, the strength distributions
associated with the Gamow-Teller operator
$\sum_{i=1}^A \vec \sigma(i) t_-(i)$ are expected to display
a main resonance located at energy $E_{\rm GTR}$. In 
Fig.~\ref{fig:trendGT} we show the behavior of 
$E_{\rm GTR}-E_{\rm IAR}$, where $E_{IAR}$ is the isobaric
analog energy, as function of $(N-Z)/A$.
Experimental data are from 
Refs.~\cite{Pham95,Akimune95,Krasznahorkay01}. 
The theoretical (Q)RPA calculations
have been performed with some of the most recent and/or
widely used Skyrme interactions, that is, 
SIII, SGII, SLy5 and SkO$^\prime$. For SIII and SGII, 
on ground of what discussed in the previous Section, the 
$J^2$ terms are included in the residual interaction and
not in the mean field. When our calculations produce 
a resonance which is fragmented in more than one peak, 
the exact definition 
of the values of $E_{\rm GTR}$ used in the figure is the centroid
$m_1/m_0$ where the two sum rules are evaluated in 
the interval of the resonance. This interval is 15-24 MeV for Pb and 
12-22 MeV for the Sn isotopes (in Zr, there is a single GT 
main state).
In some cases, we
face the well-known problem of (Q)RPA instabilities and 
(Q)TDA values are reported (in particular, this happens for
$^{90}$Zr and $^{118,120}$Sn when the force SkO$^\prime$ is
employed, and for $^{114}$Sn when using SGII). In 
Ref.~\cite{Fracasso} we have shown that our
model provides quite accurate values of $E_{\rm IAR}$ but 
in the figure, for simplicity, we have used the experimental
values for this quantity.

From Fig.~\ref{fig:trendGT}, we can draw two first conclusions.
First, one should notice that the linear behavior of 
$E_{\rm GTR}-E_{\rm IAR}$ vs. $(N-Z)/A$ was already checked, on 
the experimental data, in~\cite{Nakayama} 
as it was expected on the ground
of simple schematic models~\cite{Gaarde81,Osterfeld}.
In fact, if one performs a simple
RPA calculation using a separable interaction in a restricted space 
(made up with the excess neutrons and the proton spin-orbit
partners), one finds that
\begin{equation}
E_{\rm GTR}-E_{\rm IAR} = \Delta E_{ls} + 
2{\kappa_{\sigma\tau}-\kappa_{\tau}\over A}(N-Z),
\end{equation}
where $\Delta E_{ls}$ is (an average value of) the spin-orbit
splitting and $\kappa_{\tau}/A$ ($\kappa_{\sigma\tau}/A$) is the 
coupling constant of the separable schematic isospin (spin-isospin) 
residual force. The result of Fig.~\ref{fig:trendGT} suggests that 
our calculations, which are microscopically based and much more
sophisticated, obey in first approximation this simple pattern.

Besides that, one would also infer from the figure that 
some forces account better for the experimental findings 
while others perform less well. SkO$^\prime$ and SLy5 lie
close to experiment, although their predictions drop below
the experimental trend in $^{208}$Pb. SGII and SIII tend to
overestimate the experimental energies but the trend of SGII does  
not change abruptly for $^{208}$Pb. The result obtained
with the force SIII corresponds, within $\approx$ 400 keV, to 
the one found in Ref.~\cite{Sagawa}. The trend associated with the energy 
location of the GTR is not the only significant experimental
observable: we should also analyze the fraction of $m_0$, or
collectivity of the GTR. 

\begin{figure}[hbt]
\includegraphics[width=5.5in]{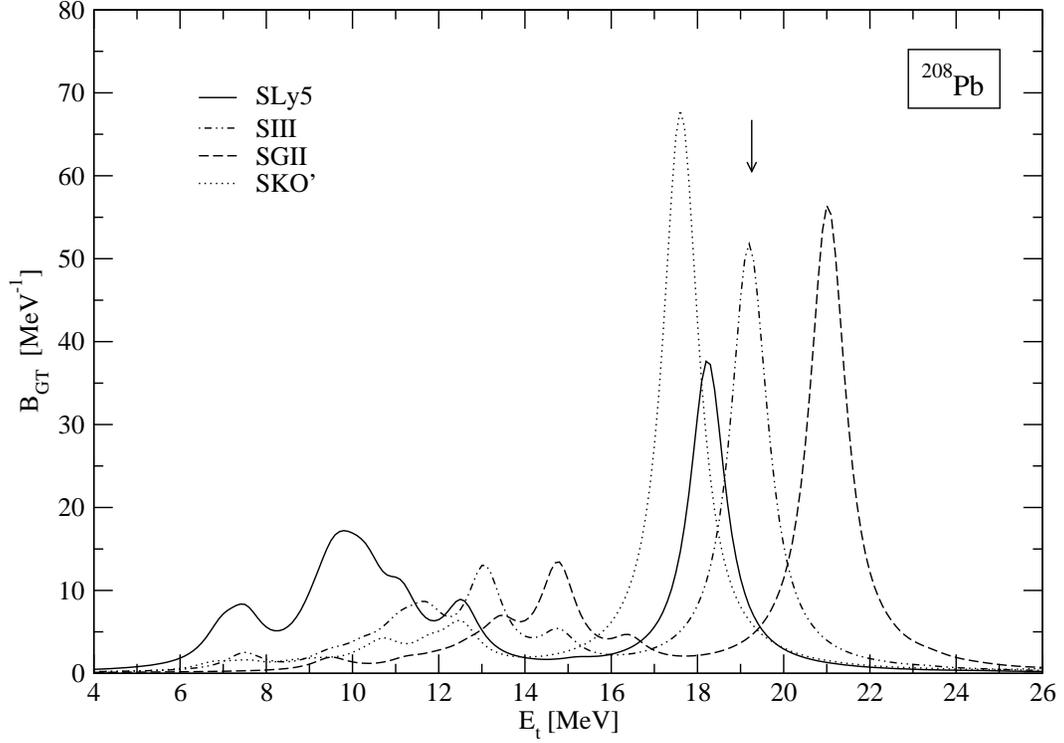}
\caption{\label{fig:BGT-Pb}
Gamow-Teller strength distributions in $^{208}$Pb, calculated
using different Skyrme forces within HF plus RPA. The results
are displayed as function of the energy in the target nucleus
($E_{\rm t}$). The discrete RPA peaks have been smeared out using 
Lorentzian functions having 1 MeV width. The arrow corresponds
to the experimental energy.
}
\end{figure}

\begin{figure}[hbt]
\includegraphics[width=5.5in]{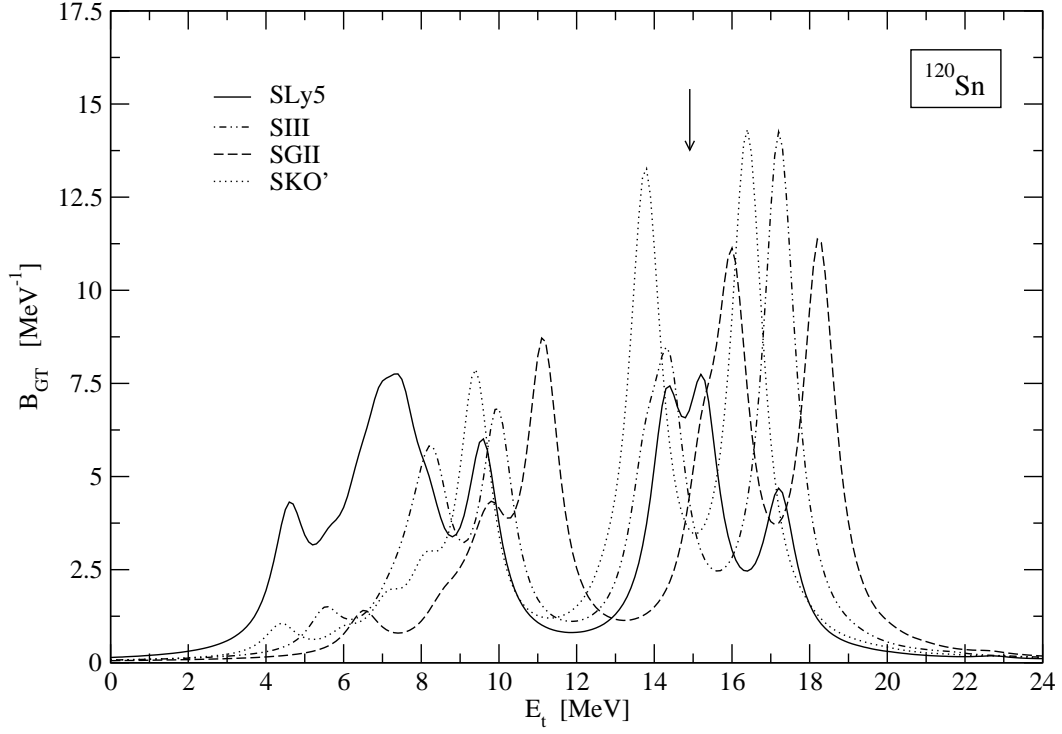}
\caption{\label{fig:BGT-Sn}
Same as Fig.~\ref{fig:BGT-Pb} for the open-shell 
nucleus $^{120}$Sn.
}
\end{figure}

In Figs.~\ref{fig:BGT-Pb} and~\ref{fig:BGT-Sn} we show
the GT strength distributions, for $^{208}$Pb and 
$^{120}$Sn, respectively, associated with
different forces. The strength functions
in Sn display more fragmentation, as expected in keeping with 
its open-shell character. As mentioned in the Introduction, 
the experimental results is that 
about 60\% of the total strength is exhausted by
the GTR. In Table~\ref{BGT} we show the fraction of 
strength in the resonance region, for the different forces, 
both in the case of $^{208}$Pb (where the result obtained with
the force SIII is very close to the 63.6\% of Ref.~\cite{Sagawa}) 
and of few selected Sn isotopes. The results present a 
clear systematics: all forces concentrate $\approx$ 
60-70\% of the strength in the resonance region, 
apart from SLy5 (we remind again that our model does not
include the coupling with 2p-2h). 

Looking at the results for the GTR associated with the different
forces, we ask ourselves if their performances depend more
on the features of the associated mean field, or rather on 
the effective interaction in the spin-isospin channel, or
on a delicate balance between the two ingredients. 
As far as the GTR energies are concerned, we did not
find clear correlations between them 
and any simple parameter. On the other hand, interesting 
correlations are found if one analyzes the GT collectivity. 
This will allow us to
draw quite strong conclusions about the Skyrme parameter
sets under study. 

In the cases of the three 
forces SIII, SGII and SkO$^\prime$ the wavefunctions
are qualitatively similar, i.e., they display a large
number of p-h components: the
wavefunction associated to the main GT state, in the case of $^{208}$Pb 
and of the force SkO$^\prime$, is reported in Table~\ref{wfGT208_SKO'}.
The wavefunction resulting from
SLy5, shown for the same nucleus in Table~\ref{wfGT208_SLy5}, 
displays instead much less
components. 
It has been checked that the reduced collectivity of
the GTR calculated using SLy5 (and characteristic not only 
of Pb but of the Sn isotopes as well) cannot be explained simply 
in terms of the differences between the unperturbed 
energies associated with this parameter set, as compared
to the other ones. Indeed, we have observed that the 
p-h matrix elements of the SLy5 force are, on the average, 
smaller than those of the other forces. 

In our analysis, we have also singled out the role of the 
velocity-dependent terms. In particular, we have observed  
that the $\left( k'^2 + k^2 \right)$ and the 
$k'k$ contributions (cf. Eq. (\ref{ph_int})) are comparable.
If we drop these terms from
the SLy5 p-h interaction, the GTR wavefunction becomes closer 
to that of the other forces, leading to an increase of the
strength of $\approx$ 20\%
(the GT energy is of course also affected).  In the case of the force SKO$^\prime$, 
the increase of collectivity when the velocity dependent terms are
dropped, is extremely small. In fact, in the 
case of SkO$^\prime$, the coefficient $C_1^{T}$, 
characterized by a positive value of $t_2$, 
is smaller as compared with the other forces. 
We conclude that both the velocity-independent and
velocity-dependent terms in the residual interaction are 
important. 

\begin{figure}[hbt]
\includegraphics[width=6.0in]{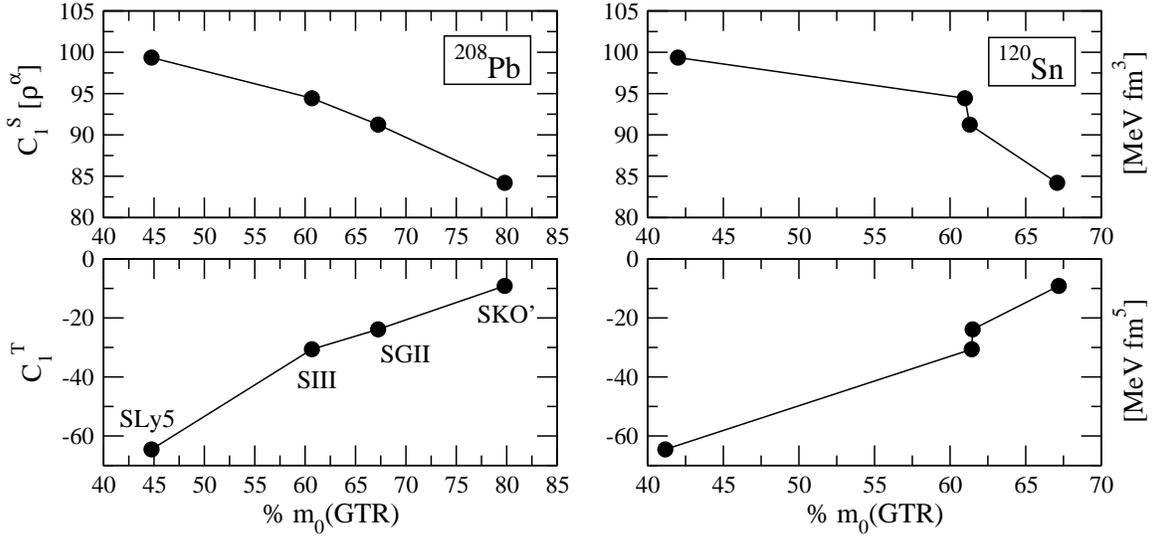}
\caption{\label{fig:Bparam}
Correlations between relevant parameters of the 
residual p-h interaction (cf. Eq. (\ref{ph_int})) 
and the percentage of $m_0$ exhausted 
by the GTR. The left (right) panel refer to $^{208}$Pb 
($^{120}$Sn). In the upper part of the figure, 
the coefficient $C_1^S$ has been evaluated at 
$\rho$=0.16 fm$^{-3}$. 
}
\end{figure}

This discussion already points out that, although 
it is not our purpose here to discuss in too much detail the 
strategy for improving the fits of effective Skyrme forces, 
we would like to strongly push forward the use of realisitic 
constraints coming from the GT properties. We
show in Fig.~\ref{fig:Bparam} direct correlations between
the percentage of $m_0$ associated with the GTR and
combinations of Skyrme parameters (actually, we find 
correlation also with the $t_1$, $t_2$ parameters 
separately and with the quantity $\Theta_S$ defined
in~\cite{Chabanat}). We are well aware that the $(t_0,t_3)$ part
of the interaction is mainly connected to the saturation 
properties of symmetric nuclear matter and the related 
value of the incompressibility, and the $t_1,t_2$ part 
must be fitted together with finite nuclei ground state 
properties. The best choice should probably be to check 
{\em a posteriori} that the value of $t_0$ and $t_3$ are
compatible with the upper panel of Fig.~\ref{fig:Bparam}, 
and impose {\em a priori} the constraint
associated with the lower panels on the $t_1,t_2$ part, 
together with the other ones which are usually imposed. 
An alternative strategy is represented by the 
possibility of fixing the odd parameter $C^T_1$ in an 
independent way with respect to the even part of the functional.

\begin{figure}[hbt]
\includegraphics[width=5.0in]{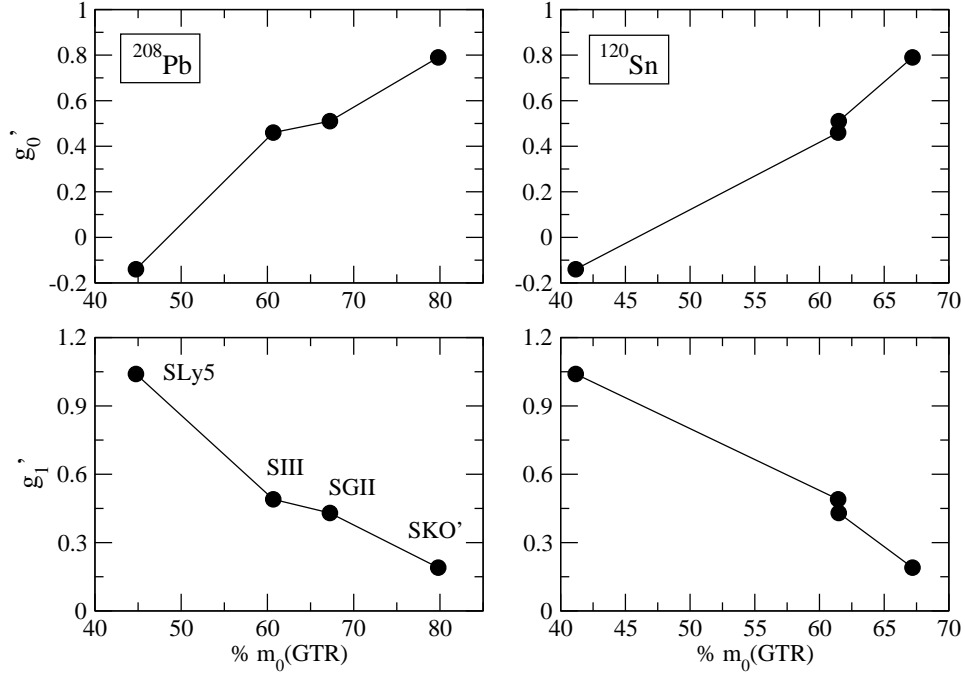}
\caption{\label{fig:Bg0}
Correlations between the Landau parameters 
$g_0^\prime$ or $g_1^\prime$ and 
the percentage of $m_0$ exhausted 
by the Gamow-Teller resonance.
The left (right) panel refer to $^{208}$Pb 
($^{120}$Sn). 
}
\end{figure}

In some works, values of the Landau parameters have been fitted.
Therefore, in Fig.~\ref{fig:Bg0} we show the
correlation between the percentage of $m_0$ exhausted 
by the GTR and either $g_0^\prime$ or $g_1^\prime$. 
Future fits of Skyrme parameter sets can certainly also benefit from
the use of one of these two costraints, which set either 
$g_0^\prime$ or $g_1^\prime$ around 0.45 or 0.5. We believe
that this estimate is more appropriate than the one
based on the empirical $g_0^\prime$ since this latter 
is, as a rule, extracted from calculations based 
on a Woods-Saxon mean field instead of a Hartree-Fock one.

In summary, our results show clearly
how the differences in the residual 
spin-isospin interaction (in particular in the 
velocity-dependent part), 
between various Skyrme parameter sets, 
manifest themselves if one studies the collectivity
of the GTR. In particular, we point to the necessity 
of new fits which include the GT data as additional
constraint, mainly to cure those forces like SLy5 which 
display a kind of anomaly in this respect. 

\begin{figure}[hbt]
\includegraphics[width=6.0in]{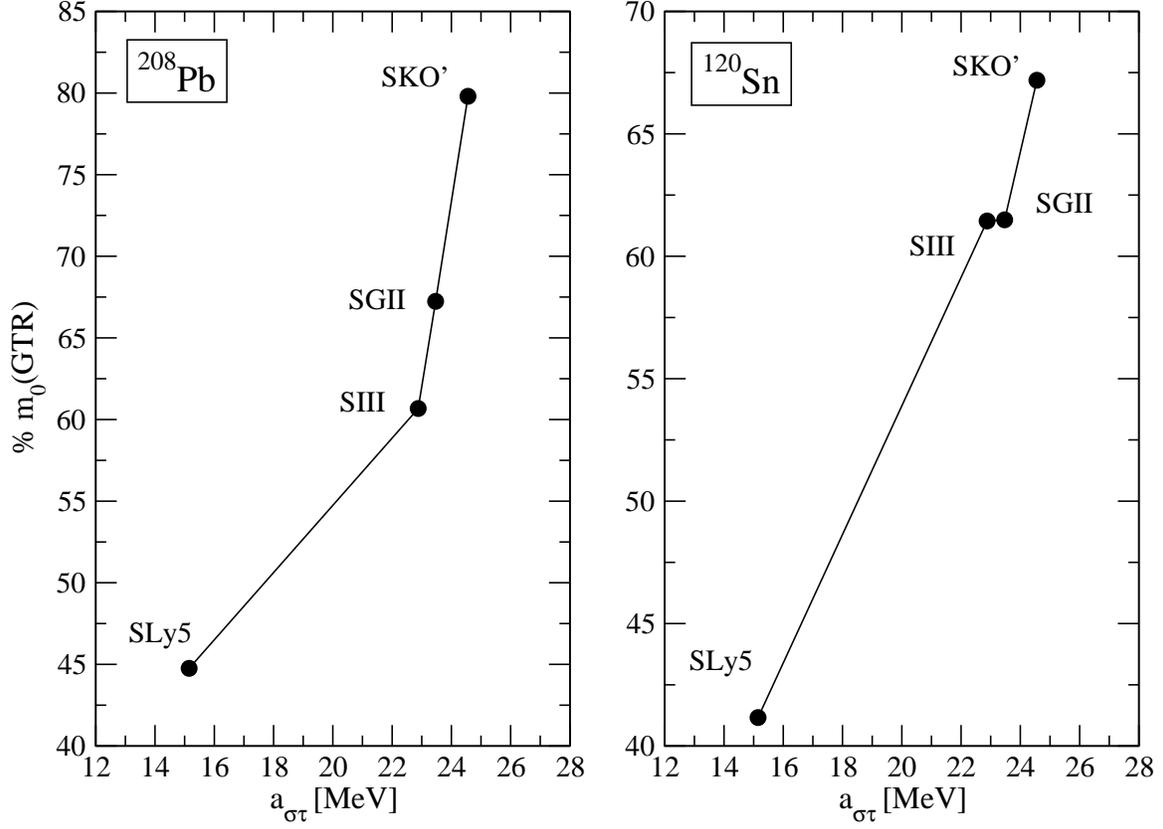}
\caption{\label{fig:Basym}
Correlation between the parameter $a_{\sigma\tau}$ 
defined in the text and the percentage of 
$m_0$ exhausted by the Gamow-Teller resonance.
The left (right) panel refer to $^{208}$Pb 
($^{120}$Sn). 
}
\end{figure}

Before concluding, we would like to show another kind
of correlation with a physical parameter 
(cf. Fig~\ref{fig:Basym}). In fact, the GT collectivity 
is also related to
the quantity which we denote by $a_{\sigma\tau}$. 
This quantity is analogous, in the spin-isospin case, 
to the well known asymmetry 
parameter $a_\tau$ (we remind that sometimes notations
like $a_4$ or $J$ are used for this latter quantity). 
It is
\begin{equation}
a_{\sigma\tau} = {1\over 2}{\partial^2\over\partial
\rho_{11}^2}
{E\over A},
\end{equation}
where we consider infinite matter with a generic spin
and isospin asymmetry, and variations
with respect to the spin-isospin density $\rho_{11}$ 
defined as
\begin{equation}
\rho_{11} = {\rho_{n\uparrow}-\rho_{n\downarrow}
-\rho_{p\uparrow}+\rho_{p\downarrow}\over\rho}.
\end{equation}

Although the spirit of our discussion is connected with the
points raised in Ref.~\cite{Bender}, our conclusions are
different. In fact, we find different results than those
published in~\cite{Bender}. We have tried to analyze in detail
the sources of this difference and in particular we have
checked the numerical effects in the case of 
$^{90}$Zr~\cite{Bender_private}. First, 
the energies in charge-exchange QRPA are naturally
defined with respect to the target nucleus ground state.
Since the experimental values of the charge-exchange
resonances are provided in the final, or daughter, systems, we 
find quite straightforward (as we did in 
the past and as other authors do) to transform the
experimental value into a corresponding value
with respect to the target nucleus ground state by
using experimental binding energies. However, this
is not done in Ref.~\cite{Bender} where a theoretical
estimate of the binding energy difference is carried 
out. In $^{90}$Zr the two alternative choices produce a
discrepancy of 1.2 MeV. A second source of difference, 
already discussed, is the treatment of the $J^2$ terms; in 
the case of $^{90}$Zr, this produce another $\approx$ 1 MeV
of difference. After considering these two facts, 
part of the discrepancy (in $^{90}$Zr, another $\approx$ 1 MeV that is
one third of the total discrepancy) has remained 
unexplained, and it is quite hard to attribute it simply to the   
different numerical implementations.

\section{Results for the spin-dipole response}
\label{IVSD_res}

The spin-dipole strength is not straightforward to 
be extracted experimentally. In absence of a well established 
proportionality between cross section at a given angle and
dipole strength, either spectra subtraction or multipole
decomposition analysis has to be attempted. On top of that,
the three different $J^\pi$ components are
mixed: the similarity of the associated angular distributions 
would require sophisticated techniques to disentangle these 
components, like the use of polarized beams or the study of 
the $\gamma$-decay of the IVSD to the GTR and to low-lying states, 
performed with high energy resolution and 
high $\gamma$-ray detection efficiency~\cite{Harakeh98}.

Theoretically, a systematic clear picture of the IVSD is
still missing. The two references mentioned in the
Introduction~\cite{Bertsch81,auerbach_spin} predict, 
respectively, the IVSD in $^{208}$Pb to lie at 21.3 and 24.0 MeV. 
Only recently self-consistent calculations have been
carried out in the same nucleus~\cite{Sagawa}, but we have 
learnt from the previous discussion on the GTR that
we need to consider several isotopes, and extract
a global trend, if we wish to understand which interactions
provide reliable results. Therefore, our present discussion
is quite timely.

\begin{figure}[hbt]
\includegraphics[width=5.5in]{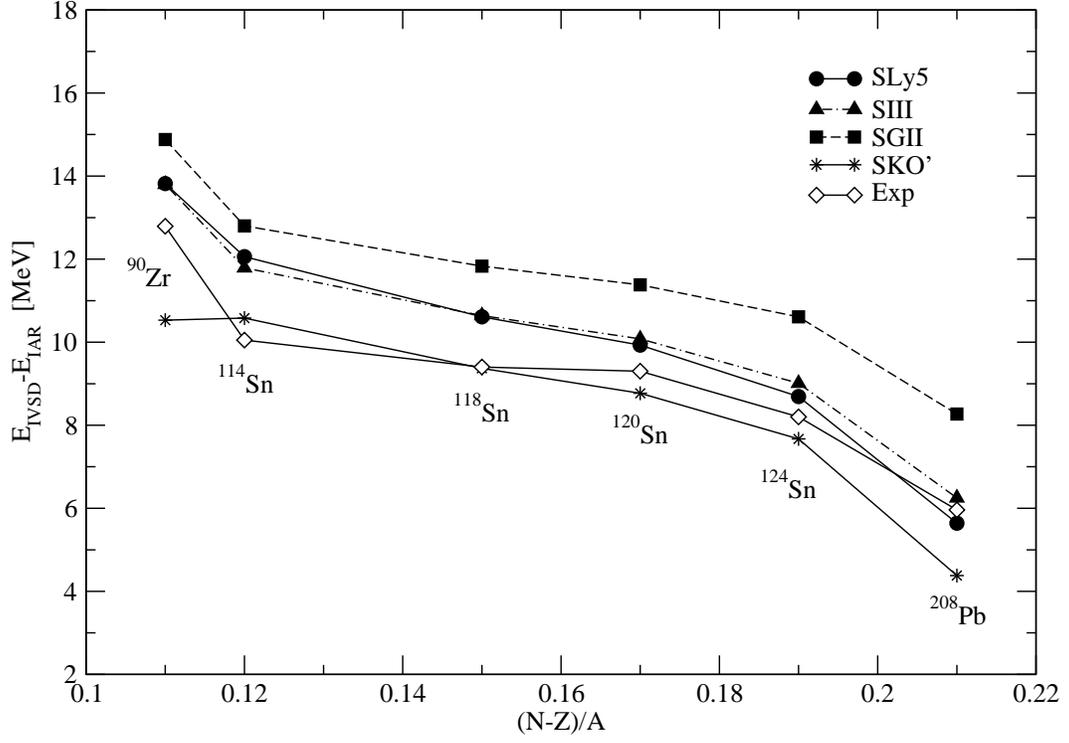}
\caption{\label{fig:trendSD}
Same as Fig.~\ref{fig:trendGT} for the spin-dipole 
case, that is, difference between the IVSD and the IAR 
energies in some selected spherical nuclei as a 
function of $(N-Z)/A$. Experimental data are 
from Refs.~\cite{Gaarde81,Krasznahorkay99,Akimune99}. 
}
\end{figure}

We of course can separate the three $J^\pi$ 
components; however, to compare with experiment,
we have to make a global average of the different 
$J^\pi$ centroids. In particular, we estimate 
\begin{equation}
\overline{E}_{IVSD_-}=\frac{\sum_{J^{\pi}=0^-,1^-,2^-}
m_1(J^{\pi})}{\sum_{J^{\pi}=0^-,1^-,2^-}m_0(J^{\pi})}, 
\label{centr}
\end{equation}
for different nuclei. We evaluate the sum rules in the 
whole energy region where the transition strength is
not negligible. We report the difference
between these energies and the IAR energies in 
Fig.~\ref{fig:trendSD} and we compare with experimental 
data from Refs.~\cite{Krasznahorkay99,Akimune99,Gaarde81}. 

\begin{figure}[hbt]
\includegraphics[width=6.0in]{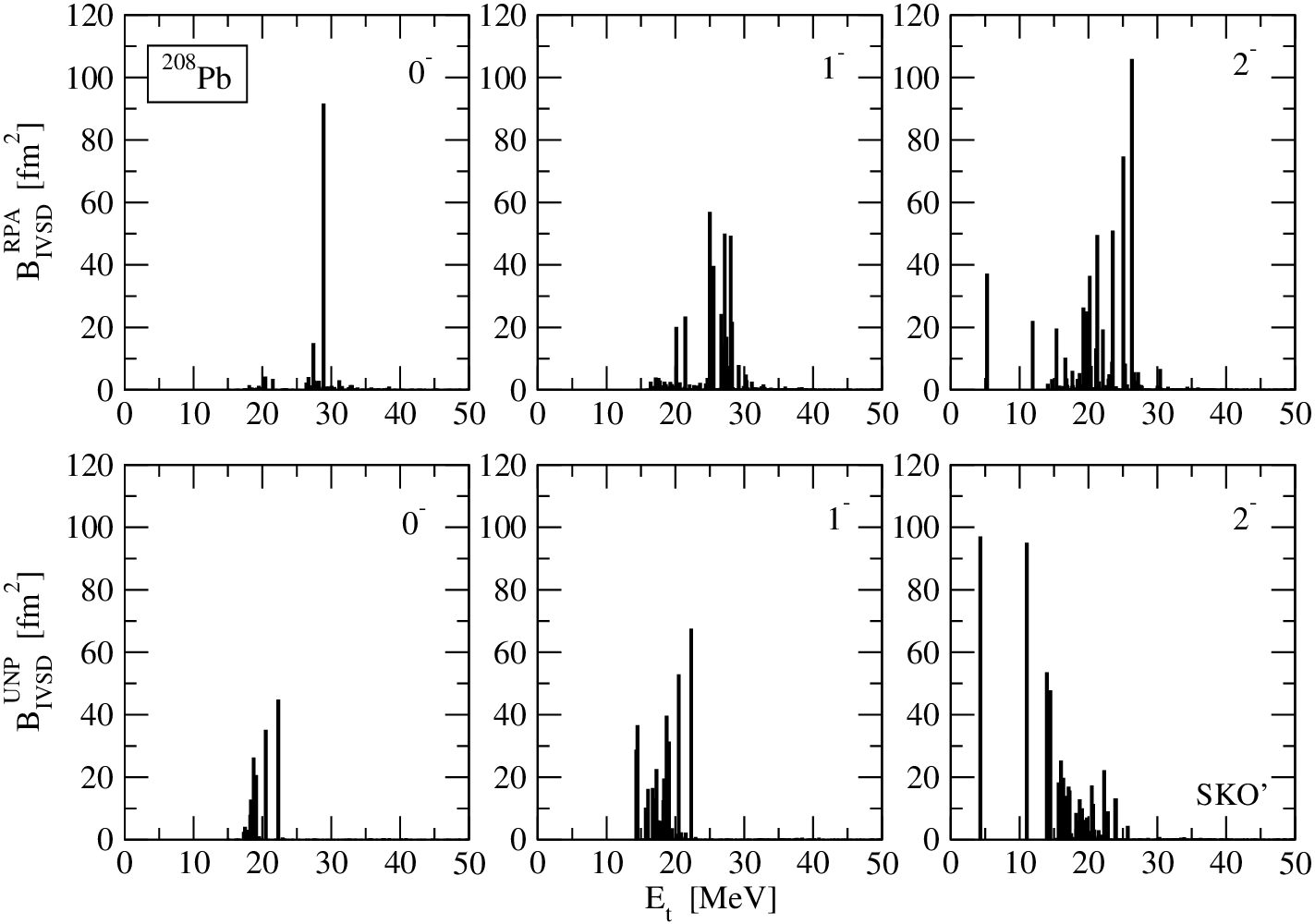}
\caption{\label{fig:IVSD_208}
RPA (upper panels) and unperturbed (lower panels) strength 
distributions for the three spin components of the IVSD. They are
calculated using the force SkO$^\prime$ in the nucleus $^{208}$Pb. 
}
\end{figure}

\begin{figure}[hbt]
\includegraphics[width=6.0in]{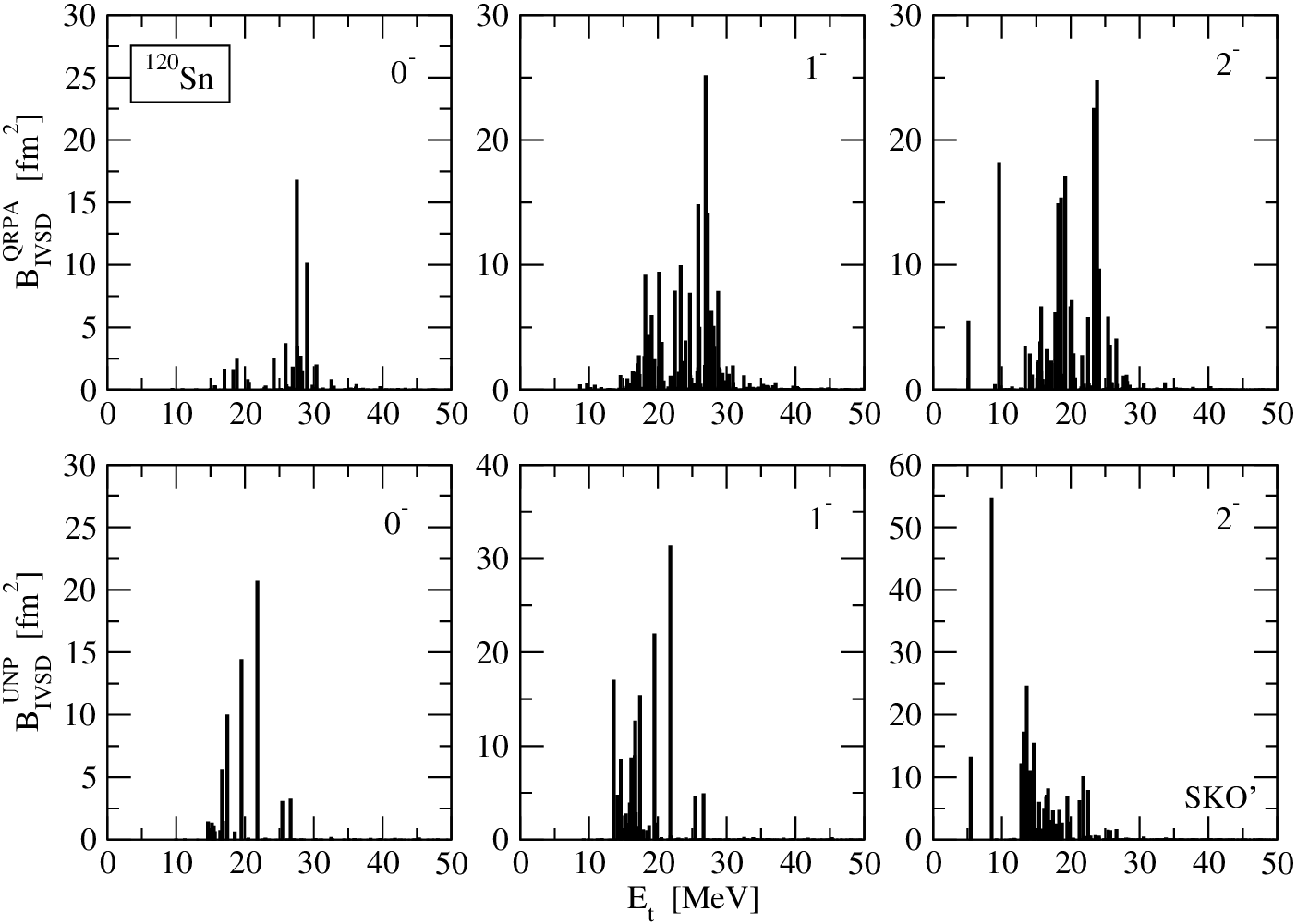}
\caption{\label{fig:IVSD_120}
QRPA (upper panels) and unperturbed (lower panels) strength 
distributions for the three spin components of the IVSD. They are
calculated using the force SkO$^\prime$ in the nucleus $^{120}$Sn. 
}
\end{figure}

\begin{figure}[hbt]
\includegraphics[width=6.0in]{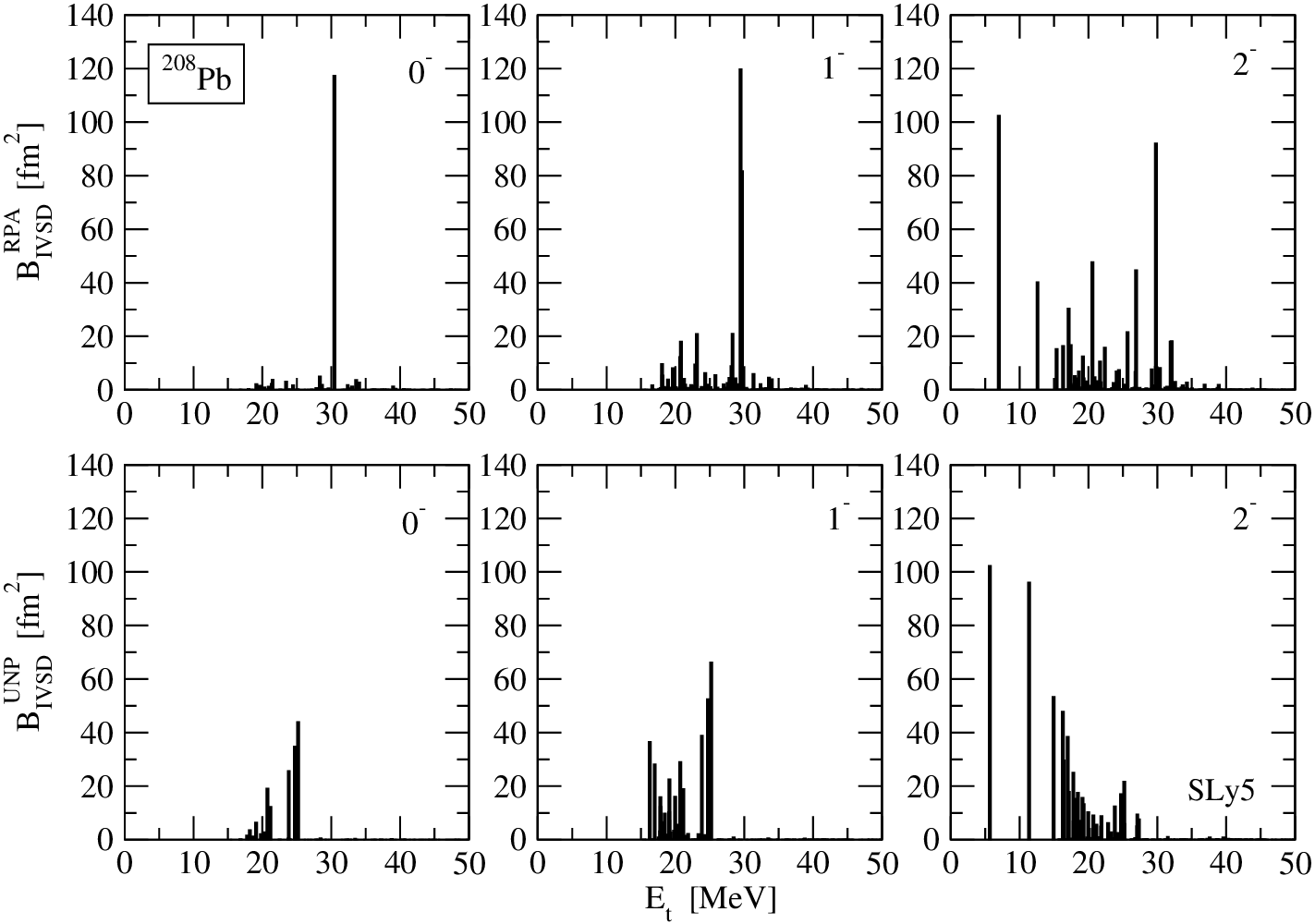}
\caption{\label{fig:IVSD_208_SLy5}
The same as Fig.~\ref{fig:IVSD_208} in the case of the 
force SLy5.}
\end{figure}

\begin{figure}[hbt]
\includegraphics[width=6.0in]{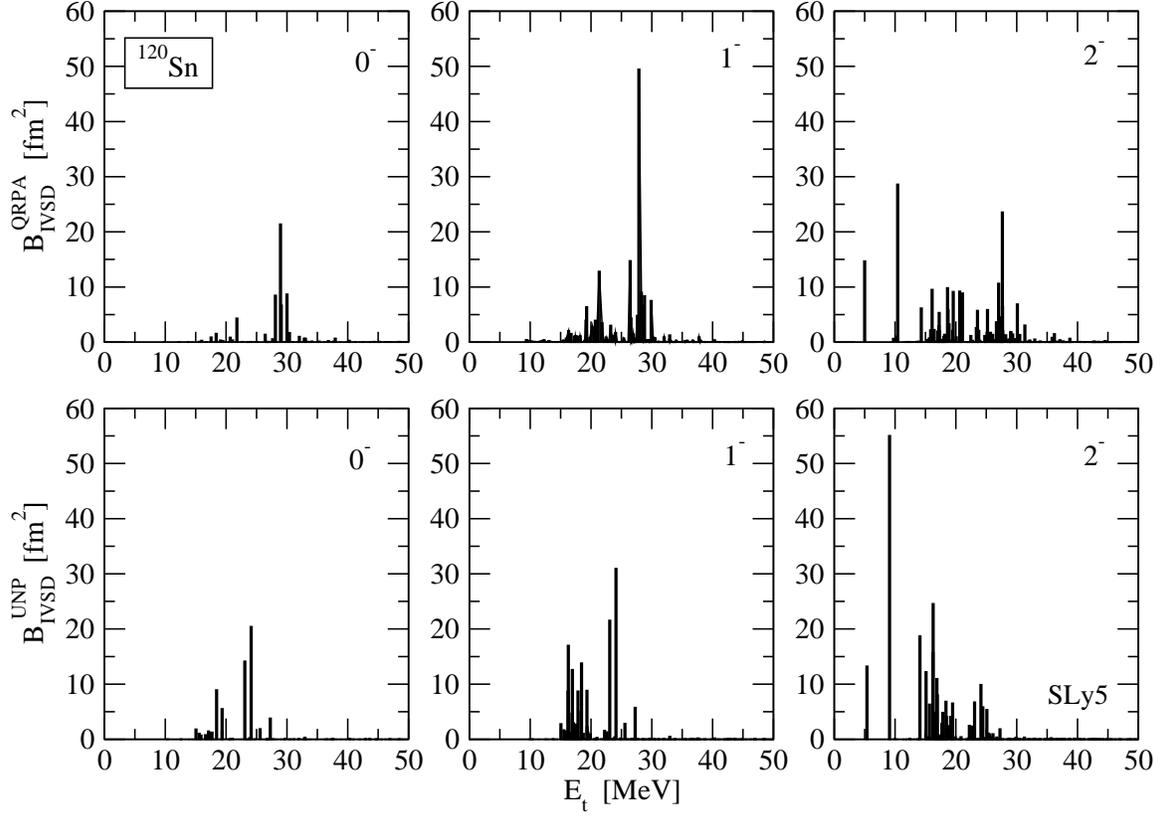}
\caption{\label{fig:IVSD_120_SLy5}
The same as Fig.~\ref{fig:IVSD_120} in the case of the 
force SLy5.}
\end{figure}

It is rather satisfactory to have found that the 
different Skyrme forces behave quite similarly, as
far as the IVSD is concerned, as they do for the simpler
GTR. We have also looked in more detail to the strength
distributions obtained by using the forces SkO$^\prime$ 
and SLy5. These distributions, for the nuclei 
$^{208}$Pb and $^{120}$Sn respectively, are displayed in 
Figs.~\ref{fig:IVSD_208} and~\ref{fig:IVSD_120} 
(SkO$^\prime$), and in Figs.~\ref{fig:IVSD_208_SLy5} 
and~\ref{fig:IVSD_120_SLy5} (SLy5). 
The complete, or (Q)RPA, strength functions are shown
in the upper panel and compared with the unperturbed
strength functions which appear in the lower panel.
The integral features of the distributions are resumed in 
Tables~\ref{DIPtab} and~\ref{DIPtab2}, for the
two forces respectively. 
It is evident that the unperturbed centroids, whose 
values are reported in parenthesis, 
follow the known energy hierarchy~\cite{Bertsch81}, the 
2$^-$ being the lowest and the 0$^-$ the highest centroid. 
This is because the 0$^-$ wavefunctions are entirely composed 
by particle-hole (or two quasiparticles) excitations between 
proton-neutron states with opposite parity and the same total 
angular momentum, which are in general widely separated in energy.
This trend is retained when the residual interaction is turned on, 
pushing up the centroids. The comparison between 
the unperturbed and the (Q)RPA distributions highlights the large values of the
repulsive matrix elements of the residual interaction.

The IVSD spectra are rather fragmented. This fragmentation
increases with the value of $L$, the 2$^-$ distribution
being broader than the 1$^-$ and 0$^-$. 
Due to the degeneracy factor, when the energy is averaged over 
the three spin-components, 
the contribution from 0$^-$ is less weighted than the 
1$^-$ and 2$^-$. 
It has been checked that the sum rules of 2$^{-}$,~1$^{-}$,~0$^{-}$ 
respect the ratio 5:~3:~1.
In the 2$^-$ spectrum of $^{208}$Pb, it 
is possible to recognize a low-lying state, due to the 
$\nu i_{13/2}\rightarrow \pi h_{9/2}$ particle-hole 
transition. Our findings are in reasonable agreement with the 
experimental peak observed, for the first time, at 
2.8 MeV (6.5 MeV referred to the target ground state) 
in Ref.~\cite{Horen80}. 

At this stage, it can be concluded that 
the behaviour of the considered Skyrme forces 
seems to be quite robust in reproducing 
properties of the isovector resonances which 
involve the spin-isospin degrees of freedom. 
Our results, reported in the figures and tables for different forces, 
can be compared with detailed forthcoming experimental 
findings (cf. e.g.~\cite{Remco}). 

\section{The effect of isovector and isoscalar pairing}
\label{pairing}

In our calculations, we are in principle sensitive to 
the effect of both isovector and isoscalar pairing.
We remind that the empirical evidence of isovector pairing, 
in the ground state of open-shell nuclei as well
as in their low-lying excitations, has been clear for long time; but, 
in connection with microscopic calculations based
on energy functionals, there is still debate about
the proper pairing force (for instance, whether it should have
volume, or surface, or mixed character). 
About isoscalar pairing, the situation is much less clear.
The existence of a $T=0$ condensate has been
questioned: if any, this 
is expected to show up only in the ground state of
nuclei having equal number of protons and neutrons, or others
lying very near. In our HF-BCS calculations, as stated
in Sec.~\ref{forma}, we fix the $T=1$ pairing force in order
to have reasonable values for the empirical pairing gaps.
The corresponding residual p-p force has been fixed by using 
the isospin invariance. If we change its strength, 
even by producing a drastic change on the pairing gap, the energy 
of the GTR is only slightly affected ($\approx$ 200 keV). 
As already said, in keeping with the lack of possible
constraints we vary the strength of the $T=0$ residual 
p-p interaction. 

In the case of the GTR, that is,
in the 1$^+$ channel, only the isoscalar residual
pairing is active when a zero-range force is assumed. 
We have studied the effects of 
the pairing correlations on the GT strength 
distributions. We have found qualitatively similar
outcomes in connection with different Skyrme forces.
In the following, we will mention some specific
results emerging from the calculations carried out
using SLy5, just for illustrative purposes: since
SLy5 does not produce highly collective GT states, 
the analysis of the effects produced by pairing is
simpler, but our general conclusions will remain
valid for other Skyrme sets.

\begin{figure}[hbt]
\includegraphics[width=5.5in]{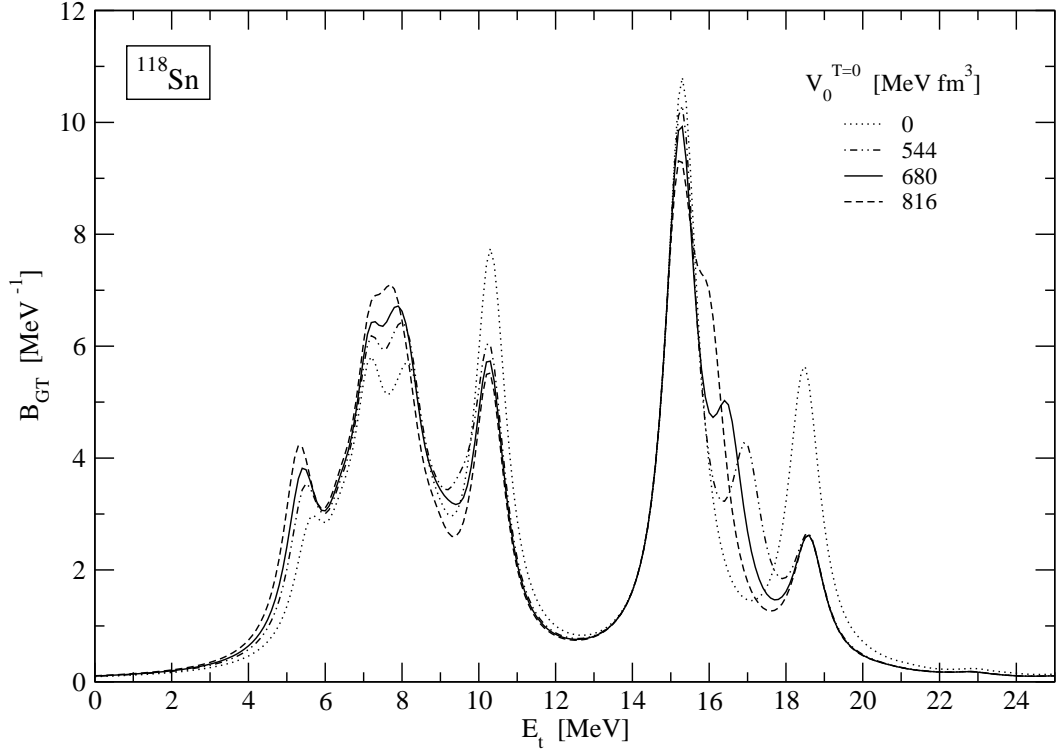}
\caption{\label{fig:GT_pairing}
GT strength distributions obtained in the nucleus $^{118}$Sn by
using the force SLy5 and varying the strength of the residual
p-p isoscalar force. 
}
\end{figure}

The effect of the residual p-p isoscalar pairing 
is shown for the isotope $^{118}$Sn in 
Fig.~\ref{fig:GT_pairing}: this effect is clearly
visible, but it is small for the main peak which
varies only by 300 keV when the pairing strength
is changed from zero to a value equal to that
of the $T=1$ pairing (i.e., 680 MeV$\cdot$fm$^3$). 
The IS pairing does not affect the total collectivity
of the GTR, leaving the considerations
made in Sec.~\ref{GT_res} basically unchanged. 

In absence of residual pairing, two peaks appear 
above 15 MeV: the first one at 15.30 MeV is mainly due to the 
$|\nu g_{9/2},\pi g_{7/2}\rangle$ configuration
while the second one at 18.47 MeV is dominated by 
the $|\nu h_{11/2},\pi h_{9/2}\rangle$ configuration. 
This so-called configuration 
splitting has been predicted~\cite{rel_paar,Guba}, 
but it is smaller than the spreading width of the GTR. 
The $|\nu h_{9/2},\pi h_{11/2}\rangle$ configuration 
gives a small QRPA solution at 18.68 MeV, which is not 
visible in the figure because of its negligible strength.

When the IS pairing is turned on, three new QRPA states 
show up, in which the mentioned configurations are
mixed (cf. Table~\ref{table:wf_ISpair}). 
The reduction of the configuration splitting 
(already remarked in~\cite{rel_paar}, and which we have
observed as a linear function of the pairing strength), and the 
mixing of spin-flip and back spin-flip configurations
associated with the $h$-orbitals, can be understood
by analyzing the matrix elements 
\begin{eqnarray}
V^{J,ph}_{p_1h'_2p'_2h_1}&=&\left<(p_1h_1)J|V_{ph}|(p'_2h'_2)J\right>
\left(u_{p_1}v_{h_1}u_{p'_2}v_{h'_2}+v_{p_1}u_{h_1}v_{p'_2}u_{h'_2}\right)
\nonumber\\
V^{J,pp}_{p_1p_2p'_1p'_2}&=&\left<(p_1p_2)J|V_{pp}|(p'_1p'_2)J\right>
\left(u_{p_1}u_{p_2}u_{p'_1}u_{p'_2}+v_{p_1}v_{p_2}v_{p'_1}v_{p'_2}\right).
\end{eqnarray}
In the case at hand, with normal proton and superfluid neutron 
components, the previous equations reduce to 
\begin{eqnarray}
V^{J,ph}_{pn'p'n}&=&\left<(pn)J|V_{ph}|(p'n')J\right>
v_{n}v_{n'}\nonumber\\
V^{J,pp}_{pnp'n'}&=&\left<(pn)J|V_{pp}|(p'n')J\right>
u_{n}u_{n'}.
\end{eqnarray}
The $|\nu g_{9/2},\pi g_{7/2}\rangle$ configuration
is not very sensitive to the isoscalar pairing, because 
the associated p-p matrix elements are weighted by 
factors which include a very small $u_n$. On the
other hand, the $u_n$ factors associated with the 
$h_{11/2}$ and $h_{9/2}$ are not so small, and the 
$\vert\nu h_{11/2},\pi h_{9/2}\rangle$ and 
$\vert\nu h_{9/2},\pi h_{11/2}\rangle$ configurations 
have p-p matrix elements larger than the 
corresponding p-h ones, which are about one half
or negligible. Therefore, 
the $\vert\nu h_{9/2},\pi h_{11/2}\rangle$ is exclusively 
admixed in the GT wavefunction by the residual (isoscalar) pairing.

By looking also at the neighboring isotope $^{120}$Sn, 
and comparing with the results obtained with the 
force SkO$^\prime$, we have reached the following
conclusion. Although the presence of the non spin-flip 
components in the GT wavefunction depends on the
p-h interaction (as discussed in Sec.~\ref{GT_res}), the 
isoscalar pairing favours this admixture. Moreover, if 
we increase the strength of the isoscalar pairing force, 
also more back spin-flip configurations (which are energetically 
less favoured) mix in the GT wavefunction. 

In summary, the effect of pairing (both $T=0$ and $T=1$ 
pairing, the latter being responsible for 
the $u$ and $v$ factors) in the resonance region 
mainly concerns the detailed microscopic structure 
of the RPA states, besides their individual strength and
energy, the GTR centroid energy being less affected 
and the associated total strength much less. 
In principle, particle decay experiments 
could shed light on the microscopic structure of the 
GTR: quantifying 
the presence of other components than the pure direct spin-flip ones in the 
GT wavefunction may highlight the features of 
corresponding pairing matrix elements. Accordingly, the
theoretical framework based on RPA plus the coupling with 
the continuum and the more complex configurations, which
has explained the proton decay from the GTR of 
$^{208}$Pb, should be extended to superfluid systems.
This is left for future work. The present study of the behavior 
of different Skyrme sets is one of the requirements before going to
more ambitious calculations. 

We have also checked the effect of isoscalar pairing on the 
IVSD. The shifts on the
$J^\pi$=0$^-$, 1$^-$ and 2$^-$ centroids, induced by 
the $T=0$ pairing with $V_0^{(T=0)}$ equal to $V_0^{(T=1)}$, 
amounts to a few hundreds of keV. In $^{118}$Sn, the total
IVSD centroid is affected by 500 keV. This effect is
not negligible but remains smaller than the variations
associated with the choice of the p-h interaction. 

\section{Conclusions}
\label{conclu}

In this work, we have tried to shed light on the systematic
behavior of the nuclear collective spin-isospin response, 
in different spherical medium-heavy nuclei, calculated
by using the microscopic Skyrme functionals. Our model
is a self-consistent QRPA based on HF-BCS, and we have
studied both the Gamow-Teller and the spin-dipole
strength distributions. We believe that the importance of
our work stems from the fact that constraining the 
microscopic functionals in the spin-isospin channel is
highly desirable if studies of exotic nuclei and 
applications for particle physics or astrophysics are
envisaged, in which the spin-isospin transitions must 
be accurately obtained. 

Pairing must be considered if the study has to be
extended to different systems for which experimental
measurements are available. The resonance properties 
depend of course mainly on the p-h interaction. We have
not only elucidated the features of the existing
functionals, but also made suggestions for future fits.
In fact, the Lyon force SLy5 does not predict the correct
GT collectivity. The other forces we have considered
more or less reproduce this collectivity (within our mean field
approximation), SGII and SIII overpredicting somehow 
the GT centroid and SkO$^\prime$ lying closer to it.
We have found a clear correlation between the GT 
collectivity and either selected combinations of
Skyrme parameters, or Landau parameters. These
correlations may be used to improve the existing
Skyrme parametrizations.

The IVSD has been systematically studied using
our microscopic QRPA. No such study is available in
the literature so far. The IVSD behavior does not
introduce new constraints but somewhat confirms 
what has been deduced from the study of the GTR.

Finally, we have also singled out the effect of
pairing (mainly its contribution to the residual
proton-neutron interaction). Its effect is not
large enough to alter the conclusions which have
been drawn concerning the interaction in the p-h
channel. However, some conclusions of this part
are also interesting. Even if pairing does not affect
so much the GT centroid and collectivity, it
induces specific admixtures in the wavefunctions. 
If experimental evidences, coming e.g. from the
particle decay, were available, we could say
that the microscopic structure of the collective
spin-isospin states may help to pin down the
features of the effective proton-neutron force
in the p-p channel, which is one of the open
questions in nuclear structure. 

\section*{Acknowledgments}

We would like to thank M. Bender for useful communications
about his work on the present subject, as well as H. Sagawa, N. Van
Giai and R. Zegers for helpful discussions.

\newpage

\section*{APPENDIX: EXPLICIT FORM OF THE RESIDUAL P-H INTERACTION}

The coefficients appearing in Eq. (\ref{ph_int}) are 
\begin{eqnarray}
C_1^{\rho}[\rho]&=&-\frac{1}{8}(t_0-2t_0x_0)
-\frac{1}{48}\rho^{\alpha}(t_3+2t_3x_3),
\nonumber\\
C_1^S[\rho]&=&-\frac{t_0}{8}-\frac{t_3}{48}\rho^{\alpha},
\nonumber\\
C_1^{\tau}&=&\frac{1}{64}(-4t_1-8t_1x_1+4t_2+8t_2x_2),
\nonumber\\
C_1^{\bigtriangleup\rho}&=&\frac{1}{64}(3t_1+6t_1x_1+t_2+2t_2x_2),
\nonumber\\
C_1^T&=&\frac{1}{16}(-t_1+t_2), 
\nonumber\\
C_1^{\bigtriangleup S}&=&\frac{1}{64}(3t_1+t_2),
\nonumber\\
C_1^{\bigtriangledown J}&=&-\frac{1}{4}W_0
\end{eqnarray}
(in the case in which the spin-orbit part of the functional
is generalized by introducing the parameters $b_4$ and 
$b_4^\prime$~\cite{PGR-HF}, 
the last expression becomes $-\frac{1}{2}b_4^\prime$). 
If we insert these expressions in (\ref{ph_int}) we find
\begin{eqnarray}
v_{01}(r)&=&\left[2\left(-\frac{t_0}{8}-\frac{1}{4}t_0x_0\right)
-\frac{1}{24}\rho^\alpha(r)\left(t_3+2t_3x_3\right)
\right]{\vec\tau}_1{\vec\tau}_2,\nonumber\\
v_{11}(r)&=&\left[-\frac{t_0}{4}-\frac{t_3}{24}\rho^\alpha(r)\right]
{\vec\sigma}_1{\vec\sigma}_2{\vec\tau}_1{\vec\tau}_2,\nonumber\\
v_{01}'&=&\left[-\frac{t_1}{8}(2x_1+1)(k'^2+k^2)+\frac{t_2}{4}(2x_2+1)(k'k)
\right]{\vec\tau}_1{\vec\tau}_2,\nonumber\\
v_{11}'&=&\left[-\frac{t_1}{8}(k'^2+k^2)+\frac{t_2}{4}k'k\right]{\vec\sigma}_1
{\vec\sigma}_2{\vec\tau}_1{\vec\tau}_2,\nonumber\\
v_{1}^{(s.o.)}&=&
{iW_0\over 2}\left(\sigma_1+\sigma_2\right)k' \times k {\vec\tau}_1{\vec\tau}_2,
\end{eqnarray}
keeping the same notation of Sec.~\ref{forma}. 

The choice of neglecting the contribution to the
residual interaction from the $J^2$ terms amounts to writing
\begin{eqnarray}
v_{01}'&=&\left[\frac{1}{16}
\left[\bigtriangledown_1\bigtriangledown_2+
\bigtriangledown_1'\bigtriangledown_2'\right]
(2x_1t_1-t_1)
+\frac{1}{4} k'k(2x_2t_2+t_2)\right]\vec\sigma_1\vec\sigma_2,\nonumber\\
v_{11}'&=&
\left[-\frac{t_1}{16}\left[\bigtriangledown_1\bigtriangledown_2+
\bigtriangledown_1'\bigtriangledown_2'\right]+\frac{t_2}{4}k'k
\right]\vec\tau_1\vec\tau_2\vec\sigma_1\vec\sigma_2.
\end{eqnarray}

As mentioned in Sec.~\ref{forma}, it is appropriate to give here
the expressions for the Landau parameters discussed in the 
paper. In symmetric nuclear matter, the $\ell$=0 and 1 spin-isospin 
parameters are
\begin{eqnarray}
g_0^\prime & = & -N_0 \left[ {1\over 4}t_0 + {1\over 24}t_3\rho^\alpha 
+ {1\over 8}k_F^2 (t_1-t_2) \right] \nonumber\\
g_1^\prime & = &  N_0 \left( {t_1\over 8} - {t_2\over 8} \right) k_F^2,
\end{eqnarray}
where $N_0=2k_Fm^*/\pi^2\hbar^2$ and $k_F$ is the Fermi momentum.
The Landau parameters are zero for $\ell > 1$. If we re-write 
the Landau parameters in terms of the coefficients of Eq. 
(\ref{ph_int}), they read
\begin{eqnarray}
g_0^\prime & = & N_0 \left( 2C_1^S + 2C_1^T k_F^2 \right), 
\nonumber\\
g_1^\prime & = & -2N_0 C_1^T k_F^2.
\end{eqnarray}

\begin{center}
\begin{table}
\begin{tabular}{cccccc}
\hline
       & $^{114}$Sn  &  $^{118}$Sn  &  $^{120}$Sn &  $^{124}$Sn & $^{208}$Pb  \\
\hline
SIII   & 60.44 & 60.98 &  61.44 & 62.76 & 60.68 \\
SGII   & 61.75 & 61.30 &  61.49 & 63.36 & 67.24 \\
SLy5   & 46.38 & 42.06 &  41.16 & 41.41 & 44.76 \\
SKO$^\prime$   & 66.06 &  67.08 & 67.19 & 72.76 & 79.80 \\
\hline
\end{tabular}
\caption{\label{BGT} Percentages of the Ikeda sum rule $m_0$
exhausted in the giant resonance region. This region is 
12-22 MeV in the Sn isotopes and 15-24 MeV in $^{208}$Pb.}
\end{table}
\end{center}

\vspace{1.5truecm}

\begin{center}
\begin{table}
\begin{tabular}{cc}
\hline
Configuration & Weight \\
\hline
$\nu$i$_{13/2}$ $\rightarrow$ $\pi$i$_{11/2}$ & 0.69 \\	
$\nu$h$_{11/2}$ $\rightarrow$ $\pi$h$_{9/2}$ & 0.49 \\	
$\nu$f$_{7/2}$ $\rightarrow$ $\pi$f$_{5/2}$ & 0.28 \\	
$\nu$i$_{13/2}$ $\rightarrow$ $\pi$i$_{13/2}$ & 0.20 \\	
$\nu$f$_{7/2}$ $\rightarrow$ $\pi$f$_{7/2}$ & 0.16 \\	
$\nu$h$_{9/2}$ $\rightarrow$ $\pi$h$_{9/2}$ & 0.16 \\	
$\nu$p$_{3/2}$ $\rightarrow$ $\pi$p$_{1/2}$ & 0.13 \\	
$\nu$p$_{3/2}$ $\rightarrow$ $\pi$p$_{3/2}$ & 0.13 \\	
$\nu$f$_{5/2}$ $\rightarrow$ $\pi$f$_{5/2}$ & 0.11 \\	
$\nu$f$_{5/2}$ $\rightarrow$ $\pi$f$_{7/2}$ & 0.15 \\	
$\nu$p$_{1/2}$ $\rightarrow$ $\pi$p$_{3/2}$ & 0.11 \\	
\hline
\end{tabular}
\caption{\label{wfGT208_SKO'} Wavefunction of the main GT
state in $^{208}$Pb obtained with the SkO$^\prime$ force. Under
the label ``weight'' we report the absolute value of the
quantity $X_{ph}+(-)^{S(J+L)}Y_{ph}$, which enter the
calculation of the B(GT) value.}
\end{table}
\end{center}

\begin{center}
\begin{table}
\begin{tabular}{cc}
\hline
Configuration & Weight \\
\hline
$\nu$i$_{13/2}$ $\rightarrow$ $\pi$i$_{11/2}$ & 0.79 \\	
$\nu$h$_{11/2}$ $\rightarrow$ $\pi$h$_{11/2}$ & 0.59 \\	
$\nu$f$_{7/2}$ $\rightarrow$ $\pi$f$_{5/2}$ & 0.11 \\	
$\nu$i$_{13/2}$ $\rightarrow$ $\pi$i$_{13/2}$ & 0.04 \\
\hline
\end{tabular}
\caption{\label{wfGT208_SLy5} Same as 
the previous Table, in the case of the SLy5 force.}
\end{table}
\end{center}

\begin{center}
\begin{table}
\begin{tabular}{cccc}
\hline
& $J^{\pi}$  &  $m_{J^{\pi}}(0)$  [fm$^2$] & $m_{J^{\pi}}(1)/m_{J^{\pi}}(0)$ [MeV] \\
\hline
           &  $0^-$ &   147.9 (162.5)     &  28.21 (20.16)   \\
$^{208}$Pb &  $1^-$ &   467.9 (436.0)     &  25.84 (18.49)   \\
           &  $2^-$ &   650.0 (667.0)     &  21.32 (14.62)   \\
           &        &                     &                  \\
           &   Tot. &   1265.8 (1265.5)   &  23.18 (16.67)   \\
\hline
           &  $0^-$ &   57.6 (65.9)       & 26.82 (20.11)    \\
$^{120}$Sn &  $1^-$ &   207.3 (179.5)     & 24.58 (18.19)    \\
           &  $2^-$ &   235.2 (256.5)     & 19.54 (14.49)    \\
           &        &                     &                  \\
           &  Tot.  &   500.2 (501.9)     &  22.47 (16.55)   \\
\hline
\end{tabular}
\caption{\label{DIPtab} (Q)RPA (HF-BCS) summed transition strengths, and 
centroid energies, for the three spin-dipole components. The total 
centroid defined by Eq. (\ref{centr}) is also reported. All results 
correspond to the SkO$^\prime$ force, as in Figs.~\ref{fig:IVSD_208} 
and~\ref{fig:IVSD_120}. 
}
\end{table}
\end{center}

\begin{center}
\begin{table}
\begin{tabular}{cccc}
\hline
& $J^{\pi}$  &  $m_{J^{\pi}}(0)$  [fm$^2$] & $m_{J^{\pi}}(1)/m_{J^{\pi}}(0)$ [MeV] \\
\hline
           &  $0^-$ &   158.8 (159.8)     &  29.84 (23.30)   \\
$^{208}$Pb &  $1^-$ &   432.7 (428.0)     &  27.21 (21.16)   \\
           &  $2^-$ &   645.8 (653.2)     &  21.25 (16.14)   \\
           &        &                     &                  \\
           &   Tot. &   1237.3 (1241.1)   &  24.44 (18.79)   \\
\hline
           &  $0^-$ &   64.8  (66.5)      & 28.31 (22.17)    \\
$^{120}$Sn &  $1^-$ &   187.7 (181.4)     & 25.72 (20.08)    \\
           &  $2^-$ &   249.7 (257.9)     & 20.83 (15.93)    \\
           &        &                     &                  \\
           &  Tot.  &   502.1 (505.9)     &  23.63 (18.24)   \\
\hline
\end{tabular}
\caption{\label{DIPtab2} The same as Table~\ref{DIPtab} in the 
case of the SLy5 force.}
\end{table}
\end{center}

\begin{center}
\begin{table}
\begin{tabular}{ccc}
\hline
Energy (percentage of $m_0$)   & Configuration & Weight    \\
\hline
                                & $\nu g_{9/2}, \pi g_{7/2}$   & 0.93       \\
15.25 MeV (26.4\%)              & $\nu d_{5/2}, \pi d_{3/2}$   & 0.07	    \\
                                & $\nu h_{11/2},\pi h_{9/2}$   & 0.31	    \\
\hline
                                & $\nu g_{9/2}, \pi g_{7/2}$   & 0.30       \\
16.49 MeV (9.7\%)               & $\nu h_{11/2},\pi h_{9/2}$   & 0.62	    \\
                                & $\nu h_{11/2},\pi h_{11/2}$  & 0.11	    \\
                                & $\nu h_{9/2},\pi h_{11/2}$   & 0.67	    \\
\hline
                                & $\nu g_{9/2}, \pi g_{7/2}$   & 0.15       \\
18.59 MeV (6.0\%)               & $\nu h_{11/2},\pi h_{9/2}$   & 0.68	    \\
                                & $\nu h_{9/2},\pi h_{11/2}$   & 0.69	    \\
\hline
\end{tabular}
\caption{Wavefunctions of the QRPA states obtained for the GTR in 
$^{118}$Sn, with the interaction SLy5, when the isoscalar residual pairing
is included and its strength is set equal to that of the isovector
pairing (namely 680 MeV$\cdot$fm$^3$).\label{table:wf_ISpair}}
\end{table}
\end{center}

\newpage

\vspace{1.0cm}

\vspace{1.0cm}

\newpage

\begin{thebibliography}{9}

\bibitem{Osterfeld} F. Osterfeld, Rev. Mod. Phys. 64, 491 (1992).

\bibitem{HarakehVanDerWoude} M.N. Harakeh and A. Van Der Woude, 
{\em Giant Resonances. Fundamental High-Frequency Modes of 
Nuclear Excitation} (Clarendon Press, Oxford, 2001). 

\bibitem{Ikeda} K. Ikeda, S. Fuji and J.I. Fujita, 
Phys. Lett. 3, 271 (1963).

\bibitem{Doering} R.R. Doering, A. Galonsky, D.R. Patterson, 
G.F. Bertsch, Phys. Rev. Lett. 35, 1691 (1975). 

\bibitem{RapaportGaarde} J. Rapaport {\em et al.}, Nucl. 
Phys. A410, 371 (1983); C. Gaarde, in {\em Proceedings of the Niels 
Bohr Centennial Conference on Nuclear Structure}, Copenhagen 
(North-Holland, Amsterdam), p. 449c. 

\bibitem{GoodmanTaddeucci} C.D. Goodman {\em et al.}, Phys. Rev. 
Lett. 44, 1755 (1980); T.N. Taddeucci {\em et al.}, Nucl. Phys. 
A469, 125 (1987).

\bibitem{2p2h} G.F. Bertsch and I. Hamamoto, Phys. Rev. C26, 1323 
(1982); S. Drozdz {\em et al.}, Phys. Lett. B189, 271 (1987).

\bibitem{Wakasa} T. Wakasa {\em et al.}, Phys. Rev. C55, 2909 (1997).

\bibitem{Col94} G. Col\`o, N. Van Giai, P.F. Bortignon, R.A. Broglia, 
Phys. Rev. C50, 1496 (1994). 

\bibitem{Dang} Nguyen Ding Dang {\em et al.}, Phys. Rev. Lett. 79, 1638 (1997).

\bibitem{Fracasso} S. Fracasso and G. Col\`o, Phys. Rev.
C72, 064310 (2005). 

\bibitem{Fracasso_previous} S. Fracasso and G. Col\`o, in 
{\em Frontiers in nuclear structure, astrophysics and reactions} 
(FINUSTAR, Kos, Greece, September 2005), S.V. Harissopulos, P.
Demetriou and R. Julin Eds. (AIP Conference Proceedings Series, vol. 831, 
Melville, New York, 2006); S. Fracasso and G. Col\`o, Yad. Fiz. 
(to be published). 

\bibitem{madrid_papers} P. Sarriguren {\em et al.}, Nucl. Phys. 
A635, 55 (1998); Nucl. Phys. A658, 13 (1999); Nucl. Phys. A691, 
631 (2001); Phys. Rev. C64, 064306 (2001). 

\bibitem{madrid_last} O. Moreno {\em et al.}, Phys. Rev. C73, 
054302 (2006). 

\bibitem{Engel} J. Engel, M. Bender, J. Dobaczewski, 
W. Nazarewicz and R. Surman, Phys. Rev. C60, 014302 (1999). 

\bibitem{Bender} M. Bender, J. Dobaczewski, J. Engel 
and W. Nazarewicz, Phys. Rev. C65, 054322 (2002). 

\bibitem{rel_arg} C. De Conti {\em et al.}, Phys. Lett. B444, 14 (1998); 
Phys. Lett. B494, 46 (2000).

\bibitem{rel_paar} N. Paar, T. Nik{\v{s}}i{\'{c}}, D. Vretenar 
and P. Ring, Phys. Rev. C69, 054303 (2004).

\bibitem{rel_ma} Z-Y. Ma, B-Q. Chen, N. Van Giai, and T. Suzuki,
Eur. Phys. J. A20, 429 (2004).

\bibitem{finelli} P. Finelli, Nucl. Phys. A (in press).

\bibitem{Kuzmin} V.A. Kuzmin and V.G. Soloviev, J. Phys. G10, 
1507 (1984). 

\bibitem{Babacan} T. Babacan, D.I. Salamov and A. K\"uc\"ukbursa, 
Phys. Rev. C71, 037303 (2005). 

\bibitem{Bertsch81} G. Bertsch, D. Cha, H. Toki, Phys. Rev. 
C24, 533 (1981).

\bibitem{auerbach_spin} N. Auerbach and A. Klein, Phys. Rev. 
C30, 1032 (1984).

\bibitem{Krasznahorkay99} A. Krasznahorkay {\em et al.}, Phys. Rev. 
Lett. 82, 3216 (1999). 

\bibitem{sag_tbp} H. Sagawa {\em et al.}, to be published; 
H. Sagawa (private communication). 

\bibitem{Beiner} M. Beiner, H. Flocard, N. Van Giai and 
Ph. Quentin, Nucl. Phys. A238, 29 (1975).

\bibitem{VanGiai} N. Van Giai and H. Sagawa, Phys. Lett. 
B106, 379 (1981). 

\bibitem{Chabanat} E. Chabanat, P. Bonche, P. Haensel, J. Meyer, 
R. Schaeffer, Nucl. Phys. A635, 231 (1998). 

\bibitem{SkOp} P.-G. Reinhard, D.J. Dean, W. Nazarewicz, 
J. Dobaczewski, J.A. Maruhn and M.R. Strayer, Phys. Rev. 
C60, 014316 (1999).

\bibitem{auerbach} N. Auerbach and A. Klein, Nucl. Phys. 
A395 (1983) 77. 

\bibitem{VautherinBrink} D. Vautherin and D.M. Brink, Phys. Rev. 
C5, 626 (1972).

\bibitem{tensor} G. Col\`o, H. Sagawa, S. Fracasso, 
P.F. Bortignon, Phys. Lett. B646 (2007) 227.

\bibitem{tensor2} J. Dobaczewski, nucl-th/0604043, D.M. Brink, 
Fl. Stancu, nucl-th/0702065.

\bibitem{Pham95} K. Pham {\em et al.}, Phys. Rev. C51, 526 (1995).

\bibitem{Akimune95} H. Akimune {\em et al.}, Phys. Rev. C52, 604 
(1995).

\bibitem{Krasznahorkay01} A. Krasznahorkay {\em et al.}, Phys. Rev. 
C64, 067302 (2001).  

\bibitem{Nakayama} K. Nakayama, A. Pio Gale$\tilde{a}$o, F. 
Krmpoti\'c, Phys. Lett. B114, 217 (1982).

\bibitem{Gaarde81} C. Gaarde, J. Rapaport, T.N. Taddeucci, 
C.D. Goodman, C.C. Foster, D.E. Bainum, C.A. Goulding, 
M.B. Greenfield, D.J. Horen, E. Sugarbaker, Nucl. Phys. A369, 
258 (1981). 

\bibitem{Sagawa} T. Suzuki and H. Sagawa, Eur. Phys. J. A9, 
49 (2000). 

\bibitem{Bender_private} M. Bender (private communication).

\bibitem{Harakeh98} M.N. Harakeh, Acta Phys. Pol. B29, 2199 (1998).

\bibitem{Akimune99} H. Akimune, I. Daito, Y. Fujita, M. Fujiwara, 
M.N. Harakeh, J. J\"anecke, M. Yosoi, Phys. Rev. C61, 011304R (1999).

\bibitem{Horen80} D.J. Horen, C.D. Goodman, C.C. Foster, C.A. Goulding, M.B.
Greenfield, J. Rapaport, D.E. Bainum, E. Sugarbaker, T.G. Masterson, 
F. Petrovich, W.G. Love, Phys. Lett. B95, 27 (1980). 

\bibitem{Remco} R. Zegers, in {\em COllective Motion in nuclei under
EXtreme conditions} (COMEX2, Sankt Goar, Germany, June 2006), Nucl.
Phys. A (in press) and references therein. 

\bibitem{Guba} V.G. Guba, M.A. Nikolaev, and M.G. Urin, 
Phys. Lett. B218, 283 (1989).

\bibitem{PGR-HF} P.-G. Reinhard, H. Flocard, Nucl. Phys. 584, 467 
(1995). 

\end{thebibliography}
\end{document}